\documentclass[prb,twocolumn,showpacs,longbibliography]{revtex4-1}
\usepackage{graphicx} 

\usepackage{amsmath,amsthm,amssymb,mathrsfs}
\usepackage{bm}

\begin{document}

\title{Critical current destabilizing perpendicular magnetization by the spin Hall effect}

\author{Tomohiro Taniguchi${}^{1}$, Seiji Mitani${}^{2}$, and Masamitsu Hayashi${}^{2}$
      }
 \affiliation{
${}^{1}$National Institute of Advanced Industrial Science and Technology (AIST), Spintronics Research Center, Tsukuba 305-8568, Japan \\
${}^{2}$National Institute for Materials Science, Tsukuba 305-0047, Japan
 }

\date{\today}
\begin{abstract}
{
The critical current needed to
destabilize the magnetization of a perpendicular ferromagnet
via the spin Hall effect is studied.
Both the dampinglike and fieldlike torques associated with the spin current generated 
by the spin Hall effect is included in the Landau-Lifshitz-Gilbert equation to model the system. 
In the absence of the fieldlike torque, 
the critical current is independent of the damping constant and is 
much larger than that of conventional spin torque switching of collinear magnetic systems, as in magnetic tunnel junctions. 
With the fieldlike torque included, 
we find that the critical current scales with the damping constant as 
$\alpha^{0}$ (i.e., damping independent),
$\alpha$, and
$\alpha^{1/2}$ depending on the sign of the fieldlike torque and other parameters such as the external field.
Numerical and analytical results show that the critical current can be significantly reduced 
when the fieldlike torque possesses the appropriate sign, 
i.e. when the effective field associated with the fieldlike torque is pointing 
opposite to the spin direction of the incoming electrons. 
These results provide a pathway to reducing the current needed to switch magnetization using the spin Hall effect. 
}
\end{abstract}


\maketitle


\section{Introduction}
\label{sec:Introduction}

The spin Hall effect \cite{dyakonov71,hirsch99,kato04} (SHE) in a nonmagnetic heavy metal
generates pure spin current flowing along
the direction perpendicular to an electric current.
The spin current excites magnetization dynamics
in a ferromagnet attached to the nonmagnetic heavy metal 
by the spin-transfer effect \cite{slonczewski96,berger96}.
There have been a number of experimental reports on
magnetization switching and steady precession induced by the spin Hall effect
\cite{yang08,ando08,liu12a,liu12b}.
These dynamics have attracted great attention recently from
the viewpoints of both fundamental physics and practical applications.


An important issue to be solved on the magnetization dynamics triggered by the spin Hall effect
is the reduction of the critical current density needed to destabilize the magnetization from its equilibrium direction,
which determines the current needed to switch the magnetization direction or to induce magnetization oscillation.
The reported critical current density for switching \cite{liu12a,pai12,yamanouchi13,fan13,cubukcu14}
or precession \cite{liu12b} is relatively high,
typically larger than $10^{7}$ A/cm${}^{2}$.
One of the reasons behind this may be related to the recently predicted damping constant independent critical current when SHE is used \cite{lee13,lee14}.
This is in contrast to spin-transfer-induced magnetization switching in a typical giant magnetoresistance (GMR) or magnetic tunnel junction (MTJ) device 
where the critical current is expected to be proportional to the Gilbert damping constant $\alpha$.
Here the magnetization dynamics is excited as a result of
the competition between the spin torque and the damping torque \cite{sun00}.
Since the damping constant for typical ferromagnet in GMR or MTJ devices is relatively small ($\alpha \sim 10^{-2}-10^{-3}$) \cite{ikeda10,iihama12},
it can explain why the critical current is larger for the SHE driven systems.
Thus in particular for device application purposes, it is crucial to find experimental conditions in which the magnetization dynamics can be excited with lower current.


Another factor that might contribute to the reduction of the critical current is the presence of the field like torque \cite{zhang02}.
In the GMR/MTJ systems,
both the conventional spin torque,
often referred to as the dampinglike torque,
and the fieldlike torque arise from the spin transfer between the conduction electrons and the magnetization \cite{slonczewski96,stiles02,zhang02,zwierzycki05,brataas06,theodonis06}.
Due to the short relaxation length of the transverse spin of the conduction electrons \cite{taniguchi08,ghosh12},
the damping like torque is typically larger than the fieldlike torque.
Indeed, the magnitude of the field like torque experimentally found in GMR/MTJ systems has been reported to be
much smaller than the damping like torque \cite{tulapurkar05,kubota08,sankey08,oh09}.
Because of its smallness,
the fieldlike torque had not been considered in estimating the critical current
in the GMR/MTJ systems \cite{sun00,grollier03,morise05,gusakova09},
although it does play a key role in particular systems \cite{taniguchi14APL,taniguchi15JAP}.
In contrast, recent experiments found that
the fieldlike torque associated with the SHE is larger than the damping like torque \cite{kim13,garello13,qiu14,pai14,kim14,torrejon14}.


The physical origin of the large SHE-induced fieldlike torque still remains unclear.
Other possible sources can be the Rashba effect \cite{miron11,kim12,garello13,haney13,haney13a},
bulk effect \cite{jamali13}, and the out of plane spin orbit torque \cite{yu14}.
Interestingly, the field like torque has been reported to show a large angular dependence \cite{garello13,qiu14,pauyac13}
(the angle between the current and the magnetization), which cannot be explained
by the conventional formalism of spin-transfer torque in GMR/MTJ systems.
The fieldlike torque acts as a torque due to an external field and modifies the energy landscape of the magnetization.
As a result, a large fieldlike torque can significantly influence the critical current.
However, the fieldlike torque had not been taken into account
in considering the current needed to destabilize the magnetization
from its equilibrium direction and thus its role is still unclear.


In this paper, we study the critical current needed to destabilize a perpendicular ferromagnet by the spin Hall effect.
The Landau-Lifshitz-Gilbert (LLG) equation with the dampinglike and fieldlike torques associated with the spin Hall effect is solved both numerically and analytically.
We find that the critical current can be significantly reduced when the fieldlike torque possesses the appropriate sign with respect to the dampinglike torque.
With the fieldlike torque included, the critical current scales with the damping constant as
$\alpha^{0}$ (i.e., damping independent),
$\alpha$, and
$\alpha^{1/2}$, depending on the sign of the fieldlike torque and other parameters.
Analytical formulas of such damping-dependent critical current are derived [Eqs. (\ref{eq:jc_LONG})-(\ref{eq:jc_beta})], 
and they show good agreement with the numerical calculations.
From these results, we find conditions in which the critical current can be significantly reduced compared to the damping-independent threshold, 
i.e., systems without the fieldlike torque.


The paper is organized as follows.
In Sec. \ref{sec:System description},
we schematically describe the system under consideration.
We discuss the definition of the critical current in
Sec. \ref{sec:Definition of critical current}.
Section \ref{sec:Numerically estimated critical current}
summarizes the dependences of the critical current
on the direction of the damping constant, the in-plane field, and the fieldlike torque
obtained by the numerical simulation.
The analytical formulas of the critical current and their comparison to the numerical simulations
are discussed in Sec. \ref{sec:Analytical formula of critical current}.
The condition at which damping-dependent critical current occurs is also discussed in this section.
The conclusion follows in Sec. \ref{sec:Conclusion}.




\section{System description}
\label{sec:System description}

\begin{figure}
\centerline{\includegraphics[width=1.0\columnwidth]{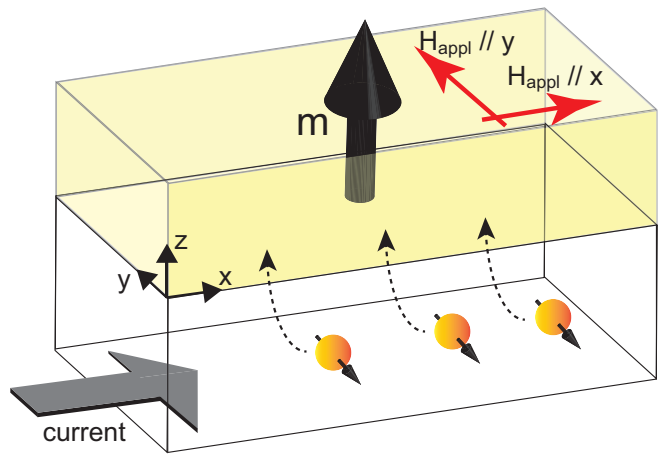}}
\caption{
         Schematic view of the spin-Hall system.
         The $x$ axis is parallel to current,
         whereas the $z$ axis is normal to the film plane.
         The spin direction of the electrons entering the magnetic layer via the spin Hall effect points along the $+y$ or $-y$ direction.
         \vspace{-3ex}}
\label{fig:fig1}
\end{figure}

The system we consider is schematically shown in Fig. \ref{fig:fig1},
where an electric current flowing along the $x$-direction
injects a spin current into the ferromagnet by the spin Hall effect.
The magnetization dynamics in the ferromagnet is
described by the LLG equation,
\begin{equation}
\begin{split}
  \frac{d \mathbf{m}}{dt}
  =&
  -\gamma
  \mathbf{m}
  \times
  \mathbf{H}
  +
  \alpha
  \mathbf{m}
  \times
  \frac{d \mathbf{m}}{dt}
\\
  &-
  \gamma
  H_{\rm s}
  \mathbf{m}
  \times
  \left(
    \mathbf{e}_{y}
    \times
    \mathbf{m}
  \right)
  -
  \gamma
  \beta
  H_{\rm s}
  \mathbf{m}
  \times
  \mathbf{e}_{y},
  \label{eq:LLG}
\end{split}
\end{equation}
where $\gamma$ and $\alpha$ are
the gyromagnetic ratio and the Gilbert damping constant, respectively.
We assume that the magnetization of the ferromagnet points along the film normal
(i.e., along the $z$ axis), and an external in-plane magnetic field is applied along the $x$ or $y$ axis.
The total magnetic field $\mathbf{H}$ is given by
\begin{equation}
  \mathbf{H}
  =
  H_{\rm appl}
  \mathbf{n}_{H}
  +
  H_{\rm K}
  m_{z}
  \mathbf{e}_{z},
  \label{eq:field}
\end{equation}
where $H_{\rm appl}$ is the external field directed along the $x$ or $y$ axis and $H_{\rm K}$ is the uniaxial anisotropy field along the $z$ axis.
$\mathbf{n}_{H}$ and $\mathbf{e}_{i}$ are unit vectors that dictate the direction of the uniaxial anisotropy field and the $i$ axis, respectively.
Here we call the external field along the $x$ and $y$ directions
the longitudinal and transverse fields, respectively.
The third and fourth terms on the right-hand side of Eq. (\ref{eq:LLG})
are the damping like and fieldlike torques associated with the spin Hall effect, respectively.
The torque strength $H_{\rm s}$ can be expressed with
the current density $j$, the spin Hall angle $\vartheta$, the saturation magnetization $M$,
and the thickness of the ferromagnet $d$, i.e.,
\begin{equation}
  H_{\rm s}
  =
  \frac{\hbar \vartheta j}{2eMd}.
  \label{eq:H_s}
\end{equation}
The ratio of the fieldlike torque to the damping like torque is
represented by $\beta$.
Recent experiments found that $\beta$ is positive and is larger than 1\cite{kim13,garello13,pai14,qiu14,kim14,torrejon14}.



The magnetization dynamics described by the LLG equation can be regarded as
a motion of a point particle on a two-dimensional energy landscape.
In the presence of the fieldlike torque,
the energy map is determined by the energy density given by \cite{taniguchi15JAP}
\begin{equation}
  \mathcal{E}
  =
  - M \int
  d \mathbf{m}
  \cdot
  \mathbf{H}
  -
  \beta M
  H_{\rm s}
  \mathbf{m}
  \cdot
  \mathbf{e}_{y}.
  \label{eq:effective_energy}
\end{equation}
Then, the external field torque and the fieldlike torque,
which are the first and fourth terms on the right-hand-side of Eq. (\ref{eq:LLG}),
can be expressed as $-\gamma \mathbf{m} \times \bm{\mathcal{B}}$,
where the effective field $\bm{\mathcal{B}}$ is
\begin{equation}
  \bm{\mathcal{B}}
  =
  -\frac{\partial \mathcal{E}}{\partial M \mathbf{m}}.
  \label{eq:effective_field}
\end{equation}
The initial state of the numerical simulation is chosen to be the direction
corresponding to the minimum of the effective energy density $\mathcal{E}$.
The explicit forms of the initial state for the longitudinal and the transverse external fields are shown in Appendix A.


We emphasize for the latter discussion in Sec. \ref{sec:Analytical formula of critical current} that,
using Eqs. (\ref{eq:LLG}), (\ref{eq:effective_energy}), and (\ref{eq:effective_field}),
the time change of the effective energy density is described as
\begin{equation}
\begin{split}
  \frac{d \mathcal{E}}{dt}
  =&
  \frac{d \mathcal{E}_{\rm s}}{dt}
  +
  \frac{d \mathcal{E}_{\alpha}}{dt}.
  \label{eq:dEdt}
\end{split}
\end{equation}
Here the first and second terms on the right-hand side are
the rates of the work done by the spin Hall torque
and the dissipation due to damping, respectively,
which are explicitly given by
\begin{equation}
  \frac{d \mathcal{E}_{\rm s}}{dt}
  =
  \gamma
  M
  H_{\rm s}
  \left[
    \mathbf{e}_{y}
    \cdot
    \bm{\mathcal{B}}
    -
    \left(
      \mathbf{m}
      \cdot
      \mathbf{e}_{y}
    \right)
    \left(
      \mathbf{m}
      \cdot
      \bm{\mathcal{B}}
    \right)
  \right],
  \label{eq:dE_s}
\end{equation}
\begin{equation}
  \frac{d \mathcal{E}_{\alpha}}{dt}
  =
  -\alpha
  \gamma
  M
  \left[
    \bm{\mathcal{B}}^{2}
    -
    \left(
      \mathbf{m}
      \cdot
      \bm{\mathcal{B}}
    \right)^{2}
  \right].
  \label{eq:dE_alpha}
\end{equation}
The sign of Eq. (\ref{eq:dE_s}) depends on
the current direction and the effective magnetic field,
while that of Eq. (\ref{eq:dE_alpha}) is always negative.


The magnetic parameters used in this paper mimic the conditions achieved in CoFeB/MgO heterostructures \cite{torrejon15PRB};
$M=1500$ emu/c.c.,
$H_{\rm K}=540$ Oe,
$\vartheta=0.1$,
$\gamma=1.76 \times 10^{7}$ rad/(Oe s),
and $d=1.0$ nm.
The value of $\beta$ is varied from $-2$, $0$, to $2$.
Note that we have used a reduced $H_{\rm K}$ (Refs. \cite{liu12,liu12a})
in order to obtain critical currents that are the same order of magnitude with that obtained experimentally. 
We confirmed that the following discussions are applicable for a large value of $H_{\rm K}(\sim 1{\rm T}$). 


\section{Definition of critical current}
\label{sec:Definition of critical current}

\begin{figure}
\centerline{\includegraphics[width=0.8\columnwidth]{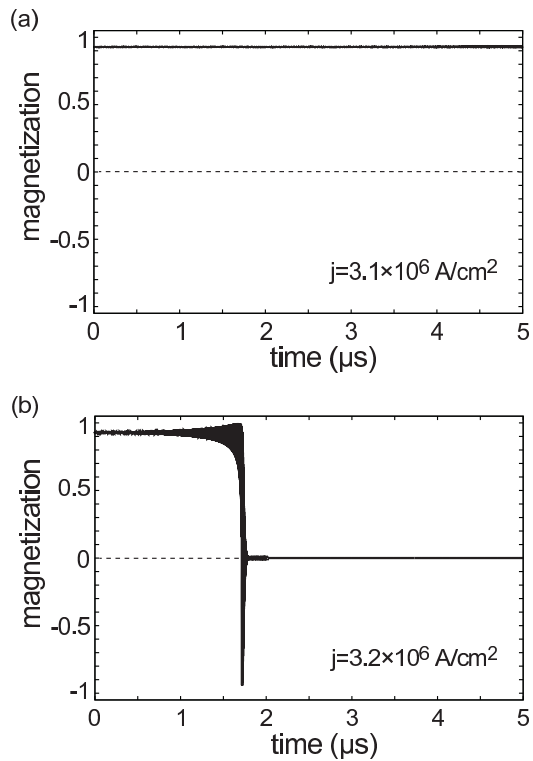}}
\caption{
         Time evolution of the $z$ component of the magnetization $m_{z}$ in the presence of the transverse field of $H_{\rm appl}=200$ with
         (a) $j=3.1 \times 10^{6}$ A/cm${}^{2}$
         and (b) $j=3.2 \times 10^{6}$ A/cm${}^{2}$.
         The value of $\beta$ is zero.
         \vspace{-3ex}}
\label{fig:fig2}
\end{figure}

In this section, we describe how we determine the critical current from the numerical simulations.
In experiments,
the critical current is determined from the observation of the magnetization reversal \cite{miron11,liu12a,liu12,fan13,torrejon15PRB,you14,yu14}.
As mentioned in Sec. \ref{sec:System description}, in this paper, 
the initial state for calculation is chosen to be the minimum of the effective energy density. 
Usually, there are two minimum points above and below the $xy$ plane because of the symmetry. 
Throughout this paper, the initial state is chosen to be the minimum point above the $xy$ plane, i.e., $m_{z}(0)>0$, for convention."
It should be noted that,
once the magnetization arrives at the $xy$ plane during the current application,
it can move to the other hemisphere after the current is turned off due to, for example, thermal fluctuation.
Therefore, here we define the critical current as the minimum current satisfying the condition
\begin{equation}
  \lim_{t \to \infty}
  m_{z}(t)
  <
  \epsilon,
  \label{eq:def_critical_current_main}
\end{equation}
where a small positive real number $\epsilon$ is chosen to be $0.001$.
The duration of the simulations is fixed to $5$ $\mu$s, long enough such that all the transient effects due to the current application are relaxed.
Figures \ref{fig:fig2}(a) and \ref{fig:fig2}(b) show examples of
the magnetization dynamics close to the critical current,
which are obtained from the numerical simulation of Eq. (\ref{eq:LLG}).
As shown, the magnetization stays near the initial state for $j=3.1 \times 10^{6}$ A/cm${}^{2}$,
while it moves to the $xy$ plane for $j=3.2 \times 10^{6}$ A/cm${}^{2}$.
Thus, the critical current is determined as $3.2 \times 10^{6}$ A/cm${}^{2}$ in this case.



We note that the choice of the definition of the critical current has some arbitrariness.
For comparison, we show numerically evaluated critical current
with a different definition in Appendix B.
The main results of this paper, e.g., the dependence of the critical current on the damping constant, are not affected by the definition.


We also point out that
the critical current defined by Eq. (\ref{eq:def_critical_current_main}) focuses on the instability threshold,
and does not guarantee a deterministic reversal.
For example, in the case of Fig. \ref{fig:fig2}(b),
the reversal becomes probabilistic
because the magnetization, starting along $+z$, stops its dynamics at the $xy$ plane
and can move back to its original direction or rotate to a point along $-z$ resulting in magnetization reversal.
Such probabilistic reversal can be measured experimentally
using transport measurements \cite{miron11,liu12,liu12a,fan13,yu14,you14} or by studying nucleation of magnetic domains via magnetic imaging \cite{torrejon15PRB}.
On the other hand, it has been reported that deterministic reversal can take place when a longitudinal in-plane field is applied alongside the current \cite{miron11,liu12}.
It is difficult to determine the critical current analytically for the deterministic switching for all conditions since, as in the case of Fig. \ref{fig:fig2}(b),
the magnetization often stops at the $xy$ plane during the current application.
This occurs especially in the presence of the transverse magnetic field
because all torques become zero at $\mathbf{m}=\pm\mathbf{e}_{y}$ and the dynamics stops.
Here we thus focus on the probabilistic reversal.





\begin{figure*}
\centerline{\includegraphics[width=2.0\columnwidth]{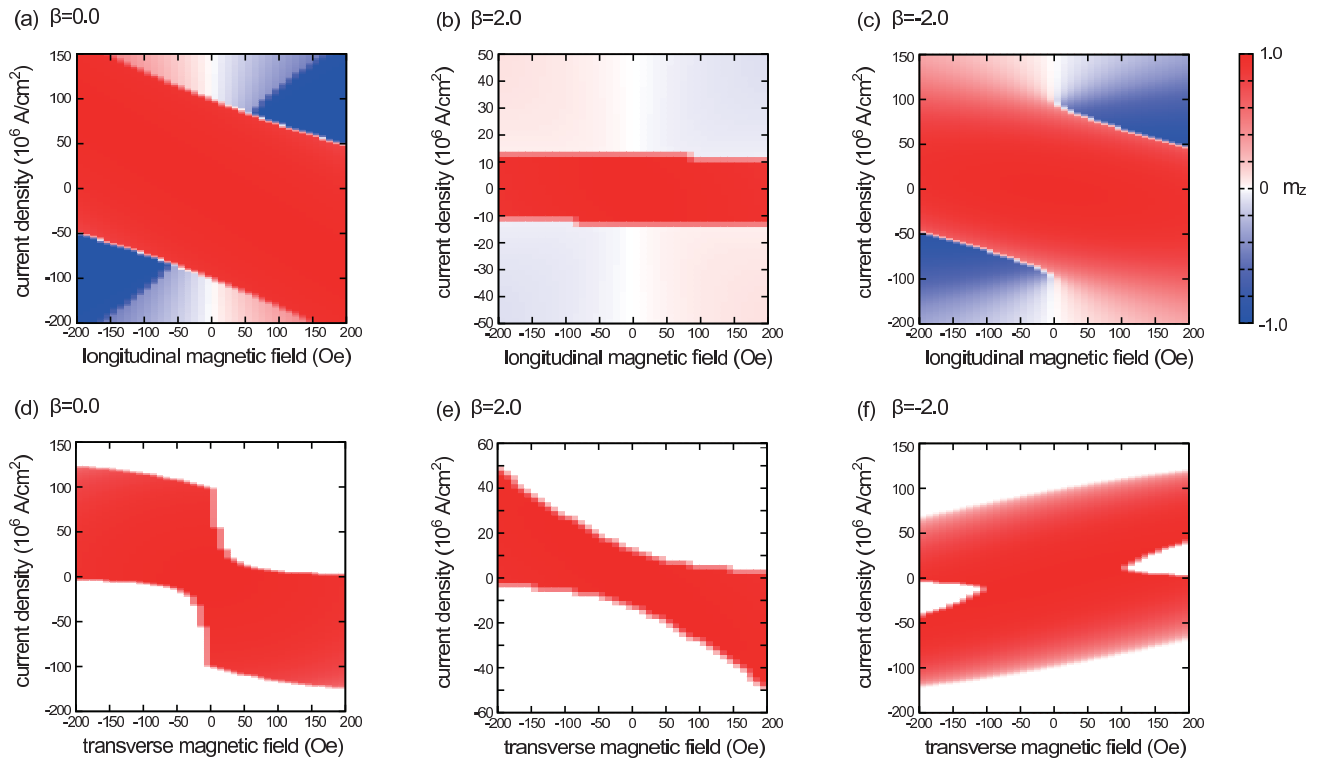}}
\caption{
         Numerically evaluated $m_{z}$ at $t=5$ $\mu$s for 
         (a)-(c) the longitudinal ($\mathbf{n}_{H}=\mathbf{e}_{x}$) and (d)-(f) the transverse ($\mathbf{n}_{H}=\mathbf{e}_{y}$) fields,
         where the value of $\beta$ is
         (a), (d) $0.0$;
         (b), (e) $2.0$;
         and (c), (f) $-2.0$.
         The damping constant is $\alpha=0.005$. 
         The color scale indicates the $z$ component of the magnetization ($m_{z}$) at $t=5$ $\mu$s.
         The red/white boundary indicates the critical current for probabilistic switching, whereas the red/blue boundary gives the critical current for deterministic switching.
         \vspace{-3ex}}
\label{fig:fig3}
\end{figure*}



\section{Numerically estimated critical current}
\label{sec:Numerically estimated critical current}

\begin{figure*}
\centerline{\includegraphics[width=2.0\columnwidth]{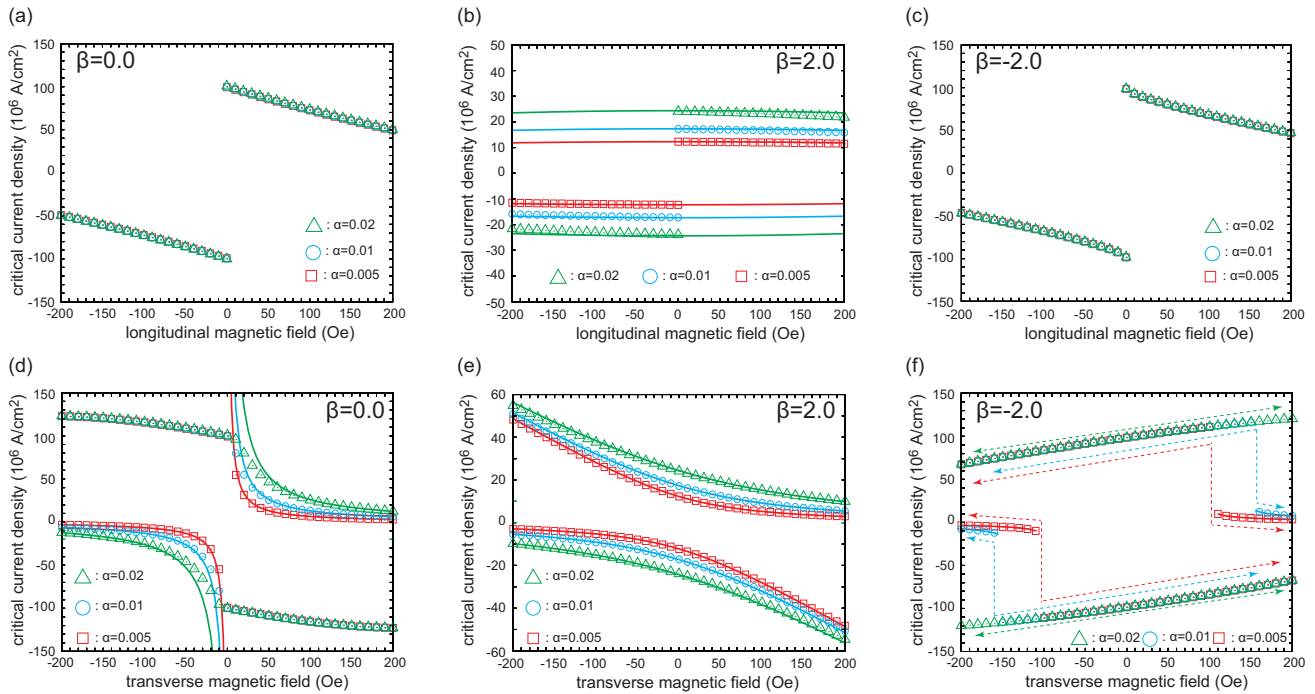}}
\caption{
         Numerically evaluated critical currents in the presence of
         (a)-(c) the longitudinal ($\mathbf{n}_{H}=\mathbf{e}_{x}$) and (d)-(f) the transverse ($\mathbf{n}_{H}=\mathbf{e}_{y}$) fields,
         where the value of $\beta$ is
         (a), (d) $0.0$;
         (b), (e) $2.0$;
         and (c), (f) $-2.0$, respectively.
         The solid lines are analytically estimated critical current in Sec. \ref{sec:Analytical formula of critical current}.
         \vspace{-3ex}}
\label{fig:fig4}
\end{figure*}

In this section, we show numerically evaluated critical current for different conditions.
We solve Eq. (\ref{eq:LLG}) and apply Eq. (\ref{eq:def_critical_current_main}) to determine the critical current. 
Figure \ref{fig:fig3} shows the value of $m_{z}$ at $t=5$ $\mu$s 
in the presence of (a)-(c) the longitudinal ($\mathbf{n}_{H}=\mathbf{e}_{x}$)
and (d)-(f) the transverse ($\mathbf{n}_{H}=\mathbf{e}_{y}$) fields.
The value of $\beta$ is $0$ for Figs. \ref{fig:fig3}(a) and \ref{fig:fig3}(d),
$2.0$ for Figs. \ref{fig:fig3}(b) and \ref{fig:fig3}(e),
and $-2.0$ for Figs. \ref{fig:fig3}(c) and \ref{fig:fig3}(f), respectively. 
The damping constant is $\alpha=0.005$. 
The red/white boundary indicates the critical current for the probabilistic switching, 
whereas the red and blue ($m_{z}=-1$) boundary gives the critical current for the deterministic switching.
Using these results and the definition of the critical current given by Eq. (\ref{eq:def_critical_current_main}),
and performing similar calculations for different values of $\alpha$,
we summarize the dependence of the critical current on the longitudinal and transverse magnetic fields in Fig. \ref{fig:fig4}.
The damping constant is varied as the following in each plot: $\alpha=0.005$, $0.01$, and $0.02$.
The solid lines in Fig. \ref{fig:fig4} represent the analytical formula derived in Sec. \ref{sec:Analytical formula of critical current}.



\subsection{In the presence of longitudinal field}

In the case of the longitudinal field and $\beta=0$ shown in Fig. \ref{fig:fig4}(a),
the critical current is damping-independent.
Such damping-independent critical current has been reported previously for deterministic magnetization switching\cite{lee13,lee14}.
Similarly, in the case of the longitudinal field and negative $\beta$ ($\beta=-2.0$) shown in Fig. \ref{fig:fig4}(c),
the critical current is damping-independent.
In these cases, the magnitude of the critical current is relatively high.
In particular, near zero field,
the critical current exceeds $\sim 10^{8}$ A/cm${}^{2}$,
which is close to the limit of experimentally accessible value.
These results indicate that the use of the longitudinal field
with zero or negative $\beta$ is ineffective for the reduction of the critical current.


On the other hand, when $\beta$ is positive,
the critical current depends on the damping constant, as shown in Fig. \ref{fig:fig4}(b).
Note that positive $\beta$ is reported for the torques associated with the spin Hall effect or Rashba effect in the heterostructures studied experimentally\cite{kim13,garello13,kim14,qiu14}.
The magnitude of the critical current, $\sim 10 \times 10^{6}$ A/cm${}^{2}$,
is relatively small compared with the cases of zero or negative $\beta$.
In this case, the use of a low damping material is effective to reduce the critical current.
Interestingly, the critical current is not proportional to the damping constant,
while that previously calculated for a GMR or MTJ system \cite{sun00} is proportional to $\alpha$.
For example,
the critical current at zero longitudinal field in Fig. \ref{fig:fig4}(b) is $12.3$, $17.2$, and $24.0$ $\times$ $10^{6}$ A/cm${}^{2}$
for $\alpha=0.005$, $0.01$, and $0.02$, respectively.
These values indicate that the critical current is proportional to $\alpha^{1/2}$.
In fact, the analytical formula derived in Sec. \ref{sec:Analytical formula of critical current}
shows that the critical current is proportional to $\alpha^{1/2}$ for positive $\beta$ 
[see Eq. (\ref{eq:jc_LONG})].


To summarize the case of the longitudinal field,
the use of a heterostructure with positive $\beta$, which is found experimentally, has the possibility to reduce the critical current if a ferromagnet with low damping constant is used.
In this case, the critical current is proportional to $\alpha^{1/2}$,
which has not been found in previous works.


\subsection{In the presence of transverse field}

In the presence of the transverse field with $\beta=0$,
the critical current shows a complex dependence on the damping constant $\alpha$,
as shown in Fig. \ref{fig:fig4}(d).
When the current and the transverse field are both positive (or negative),
the critical current is proportional to the damping constant $\alpha$ except near zero field.
The numerically calculated critical current matches well with the analytical result, Eq. (\ref{eq:jc}), shown by the solid lines.
In this case, the use of the low damping material results in the reduction of the critical current.
On the other hand, when the current and the transverse field possess the opposite sign,
the critical current is damping independent.
Moreover, in this case, the critical current is of the order of $10^{8}$ A/cm${}^{2}$.
Thus, it is preferable to use the current and field having the same sign for the reduction of the critical current.
It should be noted that, in our definition,
the same sign of current and field corresponds to the case when the direction 
of incoming electrons' spin (due to the SHE) and the transverse field are opposite to each other.
The reason why the critical current becomes damping dependent in this situation
will be explained in Sec. \ref{sec:Analytical formula of critical current}.


When $\beta$ is positive the critical current depends on the damping constant for the whole range of the transverse field, as shown in Fig. \ref{fig:fig4}(e).
The critical current is roughly proportional to $\alpha^{1/2}$, in particular, close to zero field.
The solid lines display the analytical formula, Eq. (\ref{eq:jc_beta}), and show good agreement with the numerical calculations.
The damping dependence of the critical current becomes complex when the magnitude of the transverse field is increased [see Eq. (\ref{eq:jc_beta})].
We note that the critical current for the positive $\beta$ in Fig. \ref{fig:fig4}(e) is
smaller than that for $\beta=0$ in Fig. \ref{fig:fig4}(d)
for the whole range of $H_{\rm appl}$.


On the other hand, when $\beta$ is negative,
the critical current is almost independent of $\alpha$,
especially near zero field.
However, when the transverse field is increased, there is a regime where the critical current depends on the damping constant.
Such transition of the critical current with the transverse field is also predicted by the analytical solution, Eq. (\ref{eq:jc_beta}).


To summarize the case of the transverse field,
the $\alpha$ dependence of the critical current can be categorized into the following:
$\alpha^{0}$ (damping independent),
$\alpha$,
$\alpha^{1/2}$,
or other complex behavior.
As with the case of the longitudinal field, the use of a heterostructure with positive $\beta$ allows reduction of the critical current 
when low damping ferromagnet is used.
Overall, the most efficient condition to reduce the critical current is
to use the transverse field with heterostructures that possess low $\alpha$ and positive $\beta$.
In this case, the critical current is reduced to the order of $10^{6}$ A/cm${}^{2}$.


\section{Analytical formula of critical current}
\label{sec:Analytical formula of critical current}

In this section,
we derive the analytical formula of the critical current from the linearized LLG equation \cite{comment_yan_arXiv}.
The complex dependences of the critical current on the damping constant $\alpha$ discussed in Sec. \ref{sec:Numerically estimated critical current}
are well explained by the analytical formula.
We also discuss the physical insight obtained from the analytical formulas.


\subsection{Derivation of the critical current}

To derive the critical current,
we consider the stable condition of the magnetization near its equilibrium.
It is convenient to introduce a new coordinate $XYZ$
in which the $Z$ axis is parallel to the equilibrium direction.
The rotation from the $xyz$- coordinate to the $XYZ$ coordinate is performed by the rotation matrix
\begin{equation}
  \mathsf{R}
  =
  \begin{pmatrix}
    \cos\theta & 0 & -\sin\theta \\
    0 & 1 & 0 \\
    \sin\theta & 0 & \cos\theta
  \end{pmatrix}
  \begin{pmatrix}
    \cos\varphi & \sin\varphi & 0 \\
    -\sin\varphi & \cos\varphi & 0 \\
    0 & 0 & 1
  \end{pmatrix},
  \label{eq:rotation_matrix}
\end{equation}
where $(\theta,\varphi)$ are the polar and azimuth angles of the magnetization at equilibrium.
The equilibrium magnetization direction under the longitudinal and transverse magnetic field is given by
Eqs. (\ref{eq:equilibrium_direction_x}) and (\ref{eq:equilibrium_direction_y}), respectively.
Since we are interested in small excitation of the magnetization around its equilibrium,
we assume that the components of the magnetization in the $XYZ$ coordinate satisfy
$m_{Z} \simeq 1$ and $|m_{X}|,|m_{Y}| \ll 1$.
Then, the LLG equation is linearized as
\begin{equation}
  \frac{1}{\gamma}
  \frac{d}{dt}
  \begin{pmatrix}
    m_{X} \\
    m_{Y}
  \end{pmatrix}
  +
  \mathsf{M}
  \begin{pmatrix}
    m_{X} \\
    m_{Y}
  \end{pmatrix}
  =
  -H_{\rm s}
  \begin{pmatrix}
    \cos\theta\sin\varphi \\
    \cos\varphi
  \end{pmatrix},
  \label{eq:linearized_LLG}
\end{equation}
where the components of the $2 \times 2$ matrix $\mathsf{M}$ are
\begin{equation}
  \mathsf{M}_{1,1}
  =
  \alpha
  \mathcal{B}_{X}
  -
  H_{\rm s}
  \sin\theta
  \sin\varphi,
  \label{eq:M_11}
\end{equation}
\begin{equation}
  \mathsf{M}_{1,2}
  =
  \mathcal{B}_{Y},
\end{equation}
\begin{equation}
  \mathsf{M}_{2,1}
  =
  \mathcal{B}_{X}
\end{equation}
\begin{equation}
  \mathsf{M}_{2,2}
  =
  \alpha
  \mathcal{B}_{Y}
  -
  H_{\rm s}
  \sin\theta
  \sin\varphi.
\end{equation}
Here, $\mathcal{B}_{X}$ and $\mathcal{B}_{Y}$ are defined as
\begin{equation}
  \mathcal{B}_{X}
  =
  H_{\rm appl}
  \sin\theta
  \cos(\varphi-\varphi_{H})
  +
  \beta
  H_{\rm s}
  \sin\theta
  \sin\varphi
  +
  H_{\rm K}
  \cos 2\theta,
  \label{eq:B_X}
\end{equation}
\begin{equation}
  \mathcal{B}_{Y}
  =
  H_{\rm appl}
  \sin\theta
  \cos(\varphi-\varphi_{H})
  +
  \beta
  H_{\rm s}
  \sin\theta
  \sin\varphi
  +
  H_{\rm K}
  \cos^{2}\theta,
  \label{eq:B_Y}
\end{equation}
where $\varphi_{H}$ represents the direction of the external field within the $xy$ plane: $\varphi_{H}=0$ for the longitudinal field and $\pi/2$ for the transverse field.


The solution of Eq. (\ref{eq:linearized_LLG}) is
$m_{X},m_{Y} \propto {\rm exp}\{\gamma[\pm i \sqrt{{\rm det}[\mathsf{M}]-({\rm Tr}[\mathsf{M}]/2)^{2}}-{\rm Tr}[\mathsf{M}]/2]t\}$,
where ${\det}[\mathsf{M}]$ and ${\rm Tr}[\mathsf{M}]$ are the determinant and trace of the matrix $\mathsf{M}$, respectively.
The imaginary part of the exponent determines the oscillation frequency around the $Z$ axis,
whereas the real part determines the time evolution of the oscillation amplitude.
The critical current is defined as the current at which the real part of the exponent is zero.
Then, the condition ${\rm Tr}[\mathsf{M}]=0$ gives
\begin{equation}
  \alpha
  \left(
    \mathcal{B}_{X}
    +
    \mathcal{B}_{Y}
  \right)
  -
  2 H_{\rm s}
  \sin\theta
  \sin\varphi
  =
  0,
  \label{eq:Tr=0}
\end{equation}



For the longitudinal field,
Eq. (\ref{eq:Tr=0}) gives
\begin{equation}
  j_{\rm c}^{\rm LONG}
  =
  \pm
  \frac{2 e \sqrt{\alpha} Md}{\hbar \vartheta}
  \frac{\sqrt{ 2 H_{\rm K}^{2} - H_{\rm appl}^{2}}}{\sqrt{ \beta (2 + \alpha \beta)} },
  \label{eq:jc_LONG}
\end{equation}
indicating that the critical current is roughly proportional to $\alpha^{1/2}$.
This formula works for positive $\beta$ only \cite{comment1}
if we assume $0<2+\alpha\beta \simeq 2$, which is satisfied for typical ferromagnets.
The critical current when the transverse field is applied reads
\begin{equation}
  j_{\rm c}^{\rm TRANS}
  =
  \frac{2 \alpha e Md}{\hbar \vartheta (H_{\rm appl}/H_{\rm K})}
  H_{\rm K}
  \left[
    1
    -
    \frac{1}{2}
    \left(
      \frac{H_{\rm appl}}{H_{\rm K}}
    \right)^{2}
  \right],
  \label{eq:jc}
\end{equation}
when $\beta=0$,
indicating that the critical current is proportional to $\alpha$.
The critical current for finite $\beta$ is
\begin{equation}
\begin{split}
  j_{\rm c}^{\rm TRANS}
  =&
  \frac{2 eMd}{\hbar \vartheta}
\\
  & \times
  \frac{- (1 + \alpha \beta)H_{\rm appl} \pm \sqrt{H_{\rm appl}^{2} + 2\alpha \beta (2 + \alpha \beta)H_{\rm K}^{2}} }
    {\beta ( 2 + \alpha \beta)}.
  \label{eq:jc_beta}
\end{split}
\end{equation}
Equation (\ref{eq:jc_beta}) works for the whole range of $|H_{\rm appl}|(<H_{\rm K})$
for positive $\beta$,
while it only works 
when $|H_{\rm appl}|>2 \alpha \beta(2+\alpha \beta)H_{\rm K}$ for negative $\beta$.
For example, when $\beta=-2.0$,
this condition is satisfied when $|H_{\rm appl}|>108$ Oe for $\alpha=0.005$ and $|H_{\rm appl}|>152$ Oe for $\alpha=0.01$.
However the condition is not satisfied for the present range of $H_{\rm appl}$ for $\alpha=0.02$.
The solid lines in Fig. \ref{fig:fig4}(f) show where Equation (\ref{eq:jc_beta}) is applicable.
The zero-field limits of Eqs. (\ref{eq:jc_LONG}) and (\ref{eq:jc_beta}) become identical,
\begin{equation}
  \lim_{H_{\rm appl} \to 0}
  j_{\rm c}
  =
  \pm
  \frac{2 e \sqrt{\alpha} Md}{\hbar \vartheta}
  \frac{\sqrt{2} H_{\rm K}}{\sqrt{\beta (2 + \alpha \beta)}},
\end{equation}
indicating that the critical current near zero field
is proportional to $\alpha^{1/2}$
when $\beta>0$.


\subsection{Discussions}

The solid lines in Fig. \ref{fig:fig4}(b), \ref{fig:fig4}(d), \ref{fig:fig4}(e), and \ref{fig:fig4}(f) show 
the analytical formulas, Eqs. (\ref{eq:jc_LONG}), (\ref{eq:jc}), and (\ref{eq:jc_beta}).
As evident, these formulas agree well with the numerical results
in the regions where the critical currents depend on the damping constant.
In this section,
we discuss the reason why the critical current becomes damping dependent or damping independent depending on the field direction and the sign of $\beta$.

\begin{figure}
\centerline{\includegraphics[width=0.8\columnwidth]{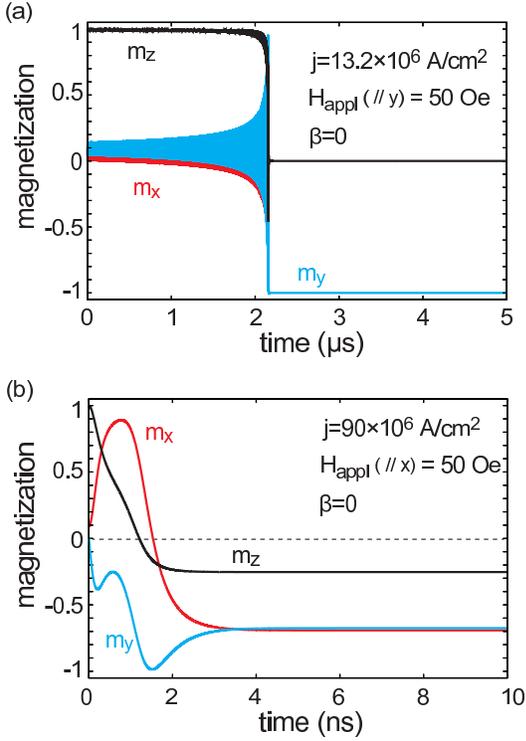}}
\caption{
         Magnetization dynamics under the conditions of
         (a) $\mathbf{n}_{H}=\mathbf{e}_{y}$, $H_{\rm appl}=50$ Oe, $\beta=0$, $\alpha=0.005$, and $j=13.2 \times 10^{6}$ A/cm${}^{2}$,
         and
         (b) $\mathbf{n}_{H}=\mathbf{e}_{x}$, $H_{\rm appl}=50$ Oe, $\beta=0$, $\alpha=0.005$, and $j=90 \times 10^{6}$ A/cm${}^{2}$.
         \vspace{-3ex}}
\label{fig:fig5}
\end{figure}

It is useful for the following discussion
to first study typical magnetization dynamics found in the numerical calculations.
Figure \ref{fig:fig5} shows the time evolution of the $x$, $y$ and $z$ components of the magnetization
when the critical current depends on [Fig. \ref{fig:fig5}(a)] or is independent of [Fig. \ref{fig:fig5}(b)] the damping constant.
For the former, the instability is accompanied with a precession of the magnetization.
On the other hand, the latter shows that the instability takes place without the precession.



\begin{figure}
\centerline{\includegraphics[width=1.0\columnwidth]{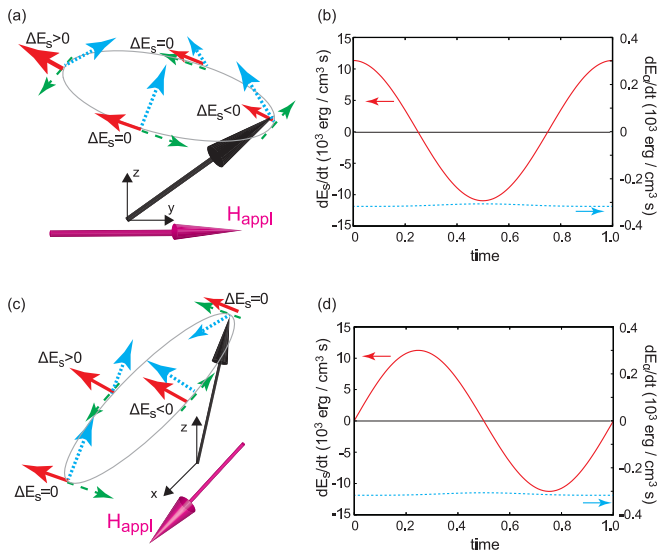}}
\caption{
         (a)
         A schematic view of the precession trajectory in the presence of the applied field in the positive $y$-direction.
         The solid and dotted arrows indicate the directions of the spin Hall torque and the damping torque, respectively.
         The dashed line, which is the tangent line to the precession trajectory, shows the field torque.
         The damping torque always dissipates energy from the ferromagnet.
         On the other hand, the spin Hall torque supplies energy ($\Delta E_{\rm s}>0$) when its direction is anti-parallel to the damping torque,
         and dissipates energy ($\Delta E_{\rm s}<0$) when the direction is parallel to the damping torque.
         When the direction of the spin Hall torque is orthogonal to the damping torque,
         the spin Hall torque does not change the energy ($\Delta E_{\rm s}=0$).
         (b)
         Typical temporal variation of the rates of the work done by the spin Hall torque, Eq. (\ref{eq:dE_s}), (solid)
         and the dissipation due to damping, Eq. (\ref{eq:dE_alpha}) (dotted) in the presence of the transverse field.
         The time is normalized by the period given by Eq. (\ref{eq:period}).
         (c), (d)
         Similar figures with the longitudinal field.
         \vspace{-3ex}}
\label{fig:fig6}
\end{figure}

We start with the case when the critical current becomes damping dependent.
To provide an intuitive picture, we schematically show in Fig. \ref{fig:fig6}(a) the torques exerted on the magnetization during one precession period when current is applied.
The condition is the same with that described in Fig. \ref{fig:fig5}(a), i.e., the transverse magnetic field is applied with $\beta=0$.
In Fig. \ref{fig:fig6}(a), magnetization is shown by the large black arrow, 
while the directions of the spin Hall torque, 
the damping torque and the external field torque are represented by the solid, dotted and dashed lines, respectively (the external field torque is tangent to the precession trajectory).
As evident in Fig. \ref{fig:fig5}(a), the precession trajectory is tilted to the positive $y$ direction due to the transverse field.
Depending on the direction of the magnetization the spin Hall torque has a component parallel, antiparallel, or normal to the damping torque.
This means that the work done by the spin Hall torque, denoted by $\Delta E_{\rm s}$ in Fig. \ref{fig:fig6} (a), is
positive, negative, or zero at these positions.
This can be confirmed numerically when we calculate the work done by the spin Hall torque using Eq. (\ref{eq:dE_s}).
For an infinitesimal time $\Delta t$, the work done by the spin Hall torque is equal to 
the rate of its work ($d \mathcal{E}_{\rm s}/dt$ ), given in Eq. (\ref{eq:dE_s}), times $\Delta t$, 
i.e. $\Delta E_{\rm s}= (d \mathcal{E}_{\rm s}/dt) \Delta t$.
The solid line in Fig. \ref{fig:fig6}(b) shows an example of the calculated rate of the work done by the spin Hall torque (solid line), $d \mathcal{E}_{\rm s}/dt$ in Eq. (\ref{eq:dE_s}).
As shown, $d \mathcal{E}_{\rm s}/dt$ is positive, negative, and zero, when the magnetization undergoes one precession period.
Similarly, the energy dissipated by the damping torque, $d \mathcal{E}_{\alpha}/dt$, can be calculated using Eq. (\ref{eq:dE_alpha}) and is shown by the dotted line in Fig. \ref{fig:fig6}(b).
The calculated dissipation due to damping over a precession period is always negative.
Details of how the rates, shown in Fig. \ref{fig:fig6}, are calculated are summarized in Appendix C.


Note that the strength of the spin Hall torque for $\Delta E_{\rm s}>0$ is larger than
that for $\Delta E_{\rm s}<0$  due to the angular dependence of the spin Hall torque, $|\mathbf{m} \times (\mathbf{e}_{y} \times \mathbf{m})|$. 
Although it is difficult to see, thesolid line in Fig. \ref{fig:fig6}(b) is slightly shifted upward. 
Thus the total energy supplied by the spin Hall torque during one precession, given by $\oint dt (d \mathcal{E}_{\rm s}/dt)$, does not average to zero and becomes positive.
When the current magnitude, $|j|$, is larger than $|j_{\rm c}|$ in Eq. (\ref{eq:jc}),
the energy supplied by the spin Hall torque overcomes the dissipation due to the damping and consequently the precession amplitude grows, 
which leads to the magnetization instability shown in Fig. \ref{fig:fig5}(a).
The same picture is applicable when both directions of field and current are reversed.
For this condition, the instability of the magnetization
is induced by the competition between the spin Hall torque and the damping torque.
Therefore, the critical current depends on the damping constant $\alpha$.
When only the current direction is reversed in Figs. \ref{fig:fig6}(a) and \ref{fig:fig6}(b) 
(i.e., the sign of the magnetic field and current is opposite to each other), 
the sign of $\Delta E_{\rm s}$ is reversed and thus the total energy supplied by the spin Hall torque becomes negative.
This means that the spin Hall torque cannot overcome the damping torque to induce instability.
Therefore, the critical current shown in Eq. (\ref{eq:jc}) only applies to the case when the sign of the field and current is the same.
As described in Sec. \ref{sec:Numerically estimated critical current}, 
the same sign of the current and field in our definition means that the incoming electrons' spin direction, 
due to the spin Hall effect, is opposite to the transverse field direction.






Next, we consider the case when the critical current is damping independent.
Figure \ref{fig:fig6} (c) schematically shows the precession trajectory
when the applied field points to the $x$ direction and $\beta=0$.
The corresponding rate of work done by the spin Hall torque and the dissipation rate due to the damping torque are shown in Fig. \ref{fig:fig6} (d).
Similar to the previous case,
$\Delta E_{\rm s}$ can be positive, negative, or zero during one precession period.
However, the total work done by the spin Hall torque, $\oint dt (d \mathcal{E}_{\rm s}/dt)$, becomes zero in this case
due to the symmetry of angular dependence of the spin Hall torque.
This means that the spin Hall torque cannot compensate the damping torque,
and thus, a steady precession assumed in the linearized LLG equation is not excited.
This is evident in the numerically calculated magnetization trajectory shown in Fig. \ref{fig:fig5}(b).
For this case, the linearized LLG equation gives $|j_{\rm c}| \to \infty$,
indicating that the spin Hall torque cannot destabilize the magnetization.
The same picture is also applicable, for example, in the absence of the applied field and $\beta=0$.


However, an alternative mechanism can cause destabilization of the magnetization.
As schematically shown in Figs. \ref{fig:fig6}(a) and \ref{fig:fig6}(c),
there is a component of the damping like spin Hall torque that is orthogonal to the damping torque when
$\Delta E_{\rm s}=0$.
The spin Hall torque at this point is parallel or antiparallel to the field torque
depending on the position of the magnetization.
When the spin Hall torque is sufficiently larger than the field torque,
the magnetization moves from its equilibrium position
even if the total energy supplied by the spin Hall torque is zero or negative.
This leads to an instability that occurs before one precession finishes.
In this case, it is expected that
the critical current is damping-independent because
the instability is induced as a competition between
the spin Hall torque and the field torque, not the damping torque.
The time evolution of the magnetization shown in Fig. \ref{fig:fig5} (b) represents such instability.
The work reported in Refs. \cite{lee13,liu12} discusses a similar instability condition.




The above physical picture is also applicable in the presence of the fieldlike torque.
The fieldlike torque, which acts like a torque due to the transverse field, modifies the equilibrium direction of the ferromagnet and thus the precession trajectory.
Consequently, the amount of energy supplied by the spin Hall torque and the dissipation due to damping is changed when the fieldlike torque is present.
Depending on the sign of $\beta$,
the amount of the work done by the spin Hall torque increases or decreases
compared to the case with $\beta=0$.
In our definition, positive $\beta$ contributes to the increase of the supplied energy,
resulting in the reduction of the critical current.
The complex dependence of the critical current on $\alpha$ arises when the fieldlike torque is present.



To summarize the discussion,
the critical current becomes damping dependent
when the energy supplied by the spin Hall torque
during a precession around the equilibrium is positive.
The condition that meets this criteria depends on the relative direction of the spin Hall torque and the damping torque, as briefly discussed above.
To derive an analytical formula that describes the condition at which the critical current becomes damping dependent is 
not an easy task except for some limited cases \cite{bertotti09}.


\section{Conclusion}
\label{sec:Conclusion}

In summary,
we have studied the critical current needed to destabilize a perpendicularly magnetized ferromagnet
by the spin Hall effect.
The Landau-Lifshitz-Gilbert (LLG) equation that includes both the dampinglike and fieldlike torques associated with the spin Hall effect is solved numerically and analytically.
The critical current is found to have different dependence on the damping constant, i.e., the critical current scales with
$\alpha^{0}$ (damping-independent),
$\alpha$, and
$\alpha^{1/2}$ depending on the sign of the fieldlike torque.
The analytical formulas of the damping-dependent critical current, Eqs. (\ref{eq:jc_LONG}), (\ref{eq:jc}), and (\ref{eq:jc_beta}),
are derived from the linearized LLG equation,
which explain well the numerical results.
We find that systems with fieldlike torque having the appropriate sign ($\beta>0$ in our definition)
are the most efficient way to reduce the critical current.
For typical material parameters found in experiment,
the critical current can be reduced to the order of $10^{6}$ A/cm${}^{2}$ when ferromagnets with reasonable parameters are used.


\section*{Acknowledgments}

The authors acknowledge T. Yorozu, Y. Shiota, and H. Kubota in AIST
for valuable discussion sthey had with us.
This work was supported by JSPS KAKENHI Grant-in-Aid for Young Scientists (B), Grant No. 25790044, 
and MEXT R \& D Next-Generation Information Technology.



\appendix


\section{Initial state of the numerical simulation}

We assume that the magnetization in the absence of the applied field points to the positive $z$ direction.
In the presence of the field, the equilibrium direction moves from the $z$ axis to the $xy$ plane.
Let us denote the zenith and azimuth angles of the initial state $\mathbf{m}(t=0)$ as
$\theta$ and $\varphi$,
i.e., $\mathbf{m}(t=0)=(\sin\theta\cos\varphi,\sin\theta\sin\varphi,\cos\theta)$.
When the applied field points to the $x$-direction ($\mathbf{n}_{H}=\mathbf{e}_{x}$),
the initial state is
\begin{equation}
  \begin{pmatrix}
    \theta \\
    \varphi
  \end{pmatrix}_{\mathbf{n}_{H}=\mathbf{e}_{x}}
  =
  \begin{pmatrix}
    \sin^{-1}[ \sqrt{H_{\rm appl}^{2} + (\beta H_{\rm s})^{2}}/H_{\rm K} ] \\
    \tan^{-1}(\beta H_{\rm s}/ H_{\rm appl})
  \end{pmatrix},
  \label{eq:equilibrium_direction_x}
\end{equation}
where the value of $\varphi$ is
$0 < \varphi < \pi/2$ for $H_{\rm appl}>0$ and $\beta H_{\rm s}>0$,
$\pi/2 < \varphi < \pi$ for $H_{\rm appl}<0$ and $\beta H_{\rm s}>0$,
$\pi < \varphi < 3\pi/2$ for $H_{\rm appl}<0$ and $\beta H_{\rm s}<0$,
and $3\pi/2 < \varphi < 2\pi$ for $H_{\rm appl}>0$ and $\beta H_{\rm s}<0$.
On the other hand,
when the applied field points to the $y$-direction ($\mathbf{n}_{H}=\mathbf{e}_{y}$),
the initial state is
\begin{equation}
  \begin{pmatrix}
    \theta \\
    \varphi
  \end{pmatrix}_{\mathbf{n}_{H}=\mathbf{e}_{y}}
  =
  \begin{pmatrix}
    \sin^{-1}[(H_{\rm appl} + \beta H_{\rm s})/H_{\rm K}] \\
    \pi/2
  \end{pmatrix},
  \label{eq:equilibrium_direction_y}
\end{equation}
where the range of the inverse sine function is $-\pi/2 \le \sin^{-1}x \le \pi/2$.
We note that the choice of the initial state does not affect the evaluation of the critical current significantly,
especially in the small field and current regimes.


\section{Numerically evaluated critical current with different definition}



\begin{figure}
\centerline{\includegraphics[width=0.8\columnwidth]{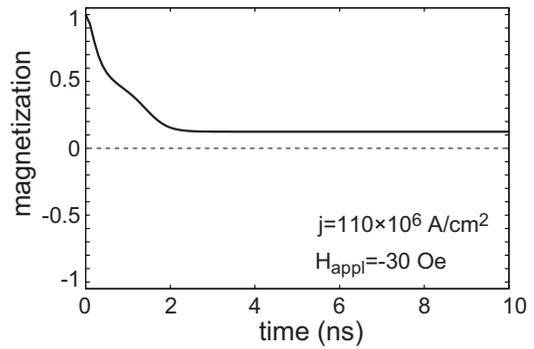}}
\caption{
         Time evolution of the $z$ component of the magnetization $m_{z}$ in the presence of the longitudinal field with
         $H_{\rm appl}=-30$ Oe, $\beta=0$, and $j=110 \times 10^{6}$ A/cm${}^{2}$.
         The dotted line is a guide showing $m_{z}=0$.
         \vspace{-3ex}}
\label{fig:fig7}
\end{figure}



As mentioned in Sec. \ref{sec:Definition of critical current},
the definition of the critical current has arbitrariness.
As an example,
we show the time evolution of $m_{z}$ under the conditions of
$\mathbf{n}_{H}=\mathbf{e}_{x}$, $H_{\rm appl}=-30$ Oe, $\beta=0$, and $j=110 \times 10^{6}$ A/cm${}^{2}$
in Fig. \ref{fig:fig7}.
In this case, the magnetization initially starts at $m_{z}=\cos[\sin^{-1}(H_{\rm appl}/H_{\rm K})]\simeq 0.99$,
and finally moves to a point $m_{z} \to 0.12$.
Since the final state does not satisfy Eq. (\ref{eq:def_critical_current_main}),
this current, $j=110 \times 10^{6}$ A/cm${}^{2}$, should be regarded as the current smaller than the critical current
in Sec. \ref{sec:Numerically estimated critical current}.
However, from the analytical point of view,
this current can be regarded as the current larger than the critical current
because the final state of the magnetization is far away from the initial equilibrium.



\begin{figure*}
\centerline{\includegraphics[width=2.0\columnwidth]{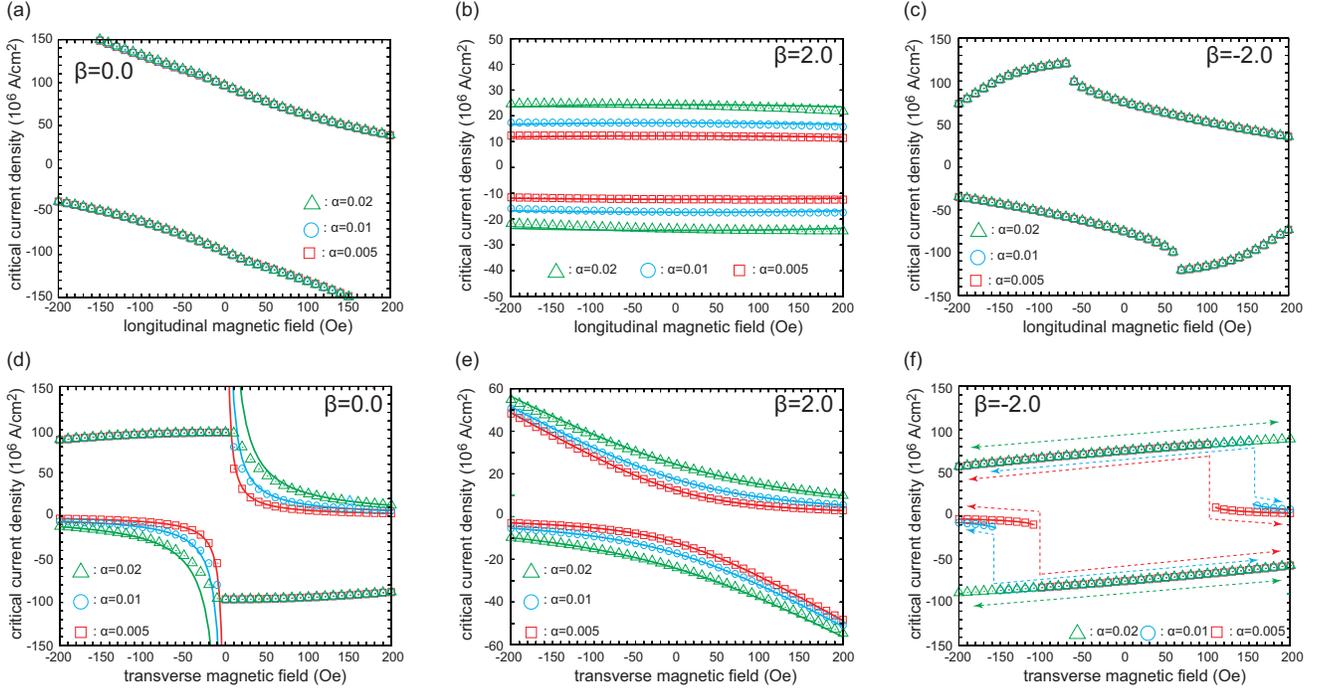}}
\caption{
         Numerically evaluated critical currents with a different definition, Eq. (\ref{eq:def_critical_current_appendix}),
         in the presence of
         (a)-(c) the longitudinal ($\mathbf{n}_{H}=\mathbf{e}_{x}$) and (d)-(f) the transverse ($\mathbf{n}_{H}=\mathbf{e}_{y}$) fields,
         where the value of $\beta$ is
         (a), (d) $0.0$;
         (b), (e) $2.0$;
         and (c), (f) $-2.0$.
         The solid lines are the analytically estimated critical current described in Sec. \ref{sec:Analytical formula of critical current}.
         \vspace{-3ex}}
\label{fig:fig8}
\end{figure*}


Regarding this point,
we show the numerically evaluated critical current with a different definition.
The magnetic state can be regarded as $unstable$
when it finally arrives at a point far away from the initial state \cite{wiggins03}.
Thus, for example, one can define the critical current as
a minimum current satisfying
\begin{equation}
  \lim_{t \to \infty}
  |m_{z}(t) - m_{z}(0)|
  >
  \delta,
  \label{eq:def_critical_current_appendix}
\end{equation}
where a small positive real number $\delta$ is chosen to be $0.1$ here.
Figure \ref{fig:fig8} summarizes the numerically evaluated
critical current with the definition of Eq. (\ref{eq:def_critical_current_appendix}).
The analytical formulas, Eqs. (\ref{eq:jc_LONG})-(\ref{eq:jc_beta}),
still fit well with the numerical results.
The absolute values of the damping-dependent critical current are slightly changed
when the definition of the critical current is changed.
This is because Eq. (\ref{eq:def_critical_current_appendix}) is more easily satisfied than Eq. (\ref{eq:def_critical_current_main}),
and thus the critical current in Fig. \ref{fig:fig8} is smaller than that shown in Fig. \ref{fig:fig4}.
However, the main results of this paper,
such as the damping dependence of the critical current, are not changed
by changing the definition of the critical current in the numerical simulations.


\section{Energy change during a precession}

As described in Sec. \ref{sec:Analytical formula of critical current},
the linearized LLG equation assumes a steady precession of the magnetization due to the field torque
when the current magnitude is close to the critical current.
This is because the spin Hall torque compensates with the damping torque.
Thus, Figs. \ref{fig:fig6}(b) and \ref{fig:fig6}(d) are obtained
by substituting the solution of $\mathbf{m}$ precessing a constant energy curve of $\mathcal{E}$
into Eqs. (\ref{eq:dE_s}) and (\ref{eq:dE_alpha}).

When the transverse field is applied and $\beta=0$, i.e., $\mathcal{E}=E$, where $E=-M \int d \mathbf{m}\cdot\mathbf{H}$, 
the precession trajectory on the constant energy curve of $E$ is given by \cite{taniguchi15}
\begin{equation}
  m_{x}(E)
  =
  (r_{2}-r_{3})
  {\rm sn}(u,k)
  {\rm cn}(u,k),
  \label{eq:mx_E}
\end{equation}
\begin{equation}
  m_{y}(E)
  =
  r_{3}
  +
  (r_{2}-r_{3})
  {\rm sn}^{2}(u,k),
  \label{eq:my_E}
\end{equation}
\begin{equation}
  m_{z}(E)
  =
  \sqrt{
    1
    -
    r_{3}^{2}
    -
    (r_{2}^{2}-r_{3}^{2})
    {\rm sn}^{2}(u,k)
  },
  \label{eq:mz_E}
\end{equation}
where $u=\gamma \sqrt{H_{\rm t} H_{\rm K}/2}\sqrt{r_{1}-r_{3}}t$, and
$r_{\ell}$ are given by
\begin{equation}
  r_{1}(E)
  =
  -\frac{E}{MH_{\rm appl}},
  \label{eq:r_1}
\end{equation}
\begin{equation}
  r_{2}(E)
  =
  \frac{H_{\rm appl}}{H_{\rm K}}
  +
  \sqrt{
    1
    +
    \left(
      \frac{H_{\rm appl}}{H_{\rm K}}
    \right)^{2}
    +
    \frac{2E}{MH_{\rm K}}
  },
  \label{eq:r_2}
\end{equation}
\begin{equation}
  r_{3}(E)
  =
  \frac{H_{\rm appl}}{H_{\rm K}}
  -
  \sqrt{
    1
    +
    \left(
      \frac{H_{\rm appl}}{H_{\rm K}}
    \right)^{2}
    +
    \frac{2E}{MH_{\rm K}}
  }.
  \label{eq:r_3}
\end{equation}
The modulus of Jacobi elliptic functions is
$k=\sqrt{(r_{2}-r_{3})/(r_{1}-r_{3})}$.
The precession period is
\begin{equation}
  \tau(E)
  =
  \frac{2 \mathsf{K}(k)}{\gamma \sqrt{H_{\rm appl} H_{\rm K}/2} \sqrt{r_{1}-r_{3}}},
  \label{eq:period}
\end{equation}
where $\mathsf{K}(k)$ is the first kind of complete elliptic integral.
The initial state is chosen to be $m_{y}(0)=r_{3}$.
Figure \ref{fig:fig6}(b) is obtained
by substituting Eqs. (\ref{eq:mx_E}), (\ref{eq:my_E}), and (\ref{eq:mz_E}) into Eqs. (\ref{eq:dE_s}) and (\ref{eq:dE_alpha}).
We note that Eqs. (\ref{eq:mx_E}), (\ref{eq:my_E}), and (\ref{eq:mz_E}) are functions of the energy density $E$.
Since we are interested in the instability threshold near the equilibrium,
the value of $E$ is chosen close to the minimum energy $E_{\rm min}$.
In Fig. \ref{fig:fig6} (b), we use $E=E_{\rm min}+(E_{\rm max}-E_{\rm min})/N$ with $N=100$,
where the minimum energy $E_{\rm min}$ and the maximum energy $E_{\rm max}$ are
$E_{\rm min}=-(M H_{\rm K}/2)[ 1 + (H_{\rm appl}/H_{\rm K})^{2}]$ and $E_{\rm max}=-MH_{\rm appl}$, respectively.
The value of $H_{\rm appl}$ is chosen to be $50$ Oe.
Figure \ref{fig:fig6}(d) is obtained in a similar way.



%


\begin{thebibliography}{55}%
\makeatletter
\providecommand \@ifxundefined [1]{%
 \@ifx{#1\undefined}
}%
\providecommand \@ifnum [1]{%
 \ifnum #1\expandafter \@firstoftwo
 \else \expandafter \@secondoftwo
 \fi
}%
\providecommand \@ifx [1]{%
 \ifx #1\expandafter \@firstoftwo
 \else \expandafter \@secondoftwo
 \fi
}%
\providecommand \natexlab [1]{#1}%
\providecommand \enquote  [1]{``#1''}%
\providecommand \bibnamefont  [1]{#1}%
\providecommand \bibfnamefont [1]{#1}%
\providecommand \citenamefont [1]{#1}%
\providecommand \href@noop [0]{\@secondoftwo}%
\providecommand \href [0]{\begingroup \@sanitize@url \@href}%
\providecommand \@href[1]{\@@startlink{#1}\@@href}%
\providecommand \@@href[1]{\endgroup#1\@@endlink}%
\providecommand \@sanitize@url [0]{\catcode `\\12\catcode `\$12\catcode
  `\&12\catcode `\#12\catcode `\^12\catcode `\_12\catcode `\%12\relax}%
\providecommand \@@startlink[1]{}%
\providecommand \@@endlink[0]{}%
\providecommand \url  [0]{\begingroup\@sanitize@url \@url }%
\providecommand \@url [1]{\endgroup\@href {#1}{\urlprefix }}%
\providecommand \urlprefix  [0]{URL }%
\providecommand \Eprint [0]{\href }%
\providecommand \doibase [0]{http://dx.doi.org/}%
\providecommand \selectlanguage [0]{\@gobble}%
\providecommand \bibinfo  [0]{\@secondoftwo}%
\providecommand \bibfield  [0]{\@secondoftwo}%
\providecommand \translation [1]{[#1]}%
\providecommand \BibitemOpen [0]{}%
\providecommand \bibitemStop [0]{}%
\providecommand \bibitemNoStop [0]{.\EOS\space}%
\providecommand \EOS [0]{\spacefactor3000\relax}%
\providecommand \BibitemShut  [1]{\csname bibitem#1\endcsname}%
\let\auto@bib@innerbib\@empty
\bibitem [{\citenamefont {Dyakonov}\ and\ \citenamefont
  {Perel}(1971)}]{dyakonov71}%
  \BibitemOpen
  \bibfield  {author} {\bibinfo {author} {\bibfnamefont {M.~I.}\ \bibnamefont
  {Dyakonov}}\ and\ \bibinfo {author} {\bibfnamefont {V.~I.}\ \bibnamefont
  {Perel}},\ }\bibfield  {title} {\enquote {\bibinfo {title} {Current-induced
  spin orientation of electrons in semiconductors},}\ }\href@noop {} {\bibfield
   {journal} {\bibinfo  {journal} {Phys. Lett. A}\ }\textbf {\bibinfo {volume}
  {35}},\ \bibinfo {pages} {459} (\bibinfo {year} {1971})}\BibitemShut
  {NoStop}%
\bibitem [{\citenamefont {Hirsch}(1999)}]{hirsch99}%
  \BibitemOpen
  \bibfield  {author} {\bibinfo {author} {\bibfnamefont {J.~E.}\ \bibnamefont
  {Hirsch}},\ }\bibfield  {title} {\enquote {\bibinfo {title} {Spin {Hall}
  {Effect}},}\ }\href@noop {} {\bibfield  {journal} {\bibinfo  {journal} {Phys.
  Rev. Lett.}\ }\textbf {\bibinfo {volume} {83}},\ \bibinfo {pages} {1834}
  (\bibinfo {year} {1999})}\BibitemShut {NoStop}%
\bibitem [{\citenamefont {Kato}\ \emph {et~al.}(2004)\citenamefont {Kato},
  \citenamefont {Myers}, \citenamefont {Gossard},\ and\ \citenamefont
  {Awschalom}}]{kato04}%
  \BibitemOpen
  \bibfield  {author} {\bibinfo {author} {\bibfnamefont {Y.~K.}\ \bibnamefont
  {Kato}}, \bibinfo {author} {\bibfnamefont {R.~C.}\ \bibnamefont {Myers}},
  \bibinfo {author} {\bibfnamefont {A.~C.}\ \bibnamefont {Gossard}}, \ and\
  \bibinfo {author} {\bibfnamefont {D.~D.}\ \bibnamefont {Awschalom}},\
  }\bibfield  {title} {\enquote {\bibinfo {title} {Observation of the {Spin}
  {Hall} {Effect} in {Semiconductors}},}\ }\href@noop {} {\bibfield  {journal}
  {\bibinfo  {journal} {Science}\ }\textbf {\bibinfo {volume} {306}},\ \bibinfo
  {pages} {1910} (\bibinfo {year} {2004})}\BibitemShut {NoStop}%
\bibitem [{\citenamefont {Slonczewski}(1996)}]{slonczewski96}%
  \BibitemOpen
  \bibfield  {author} {\bibinfo {author} {\bibfnamefont {J.~C.}\ \bibnamefont
  {Slonczewski}},\ }\bibfield  {title} {\enquote {\bibinfo {title}
  {Current-driven excitation of magnetic multilayers},}\ }\href@noop {}
  {\bibfield  {journal} {\bibinfo  {journal} {J. Magn. Magn. Mater.}\ }\textbf
  {\bibinfo {volume} {159}},\ \bibinfo {pages} {L1} (\bibinfo {year}
  {1996})}\BibitemShut {NoStop}%
\bibitem [{\citenamefont {Berger}(1996)}]{berger96}%
  \BibitemOpen
  \bibfield  {author} {\bibinfo {author} {\bibfnamefont {L.}~\bibnamefont
  {Berger}},\ }\bibfield  {title} {\enquote {\bibinfo {title} {Emission of spin
  waves by a magnetic multilayer traversed by a current},}\ }\href@noop {}
  {\bibfield  {journal} {\bibinfo  {journal} {Phys. Rev. B}\ }\textbf {\bibinfo
  {volume} {54}},\ \bibinfo {pages} {9353} (\bibinfo {year}
  {1996})}\BibitemShut {NoStop}%
\bibitem [{\citenamefont {Yang}\ \emph {et~al.}(2008)\citenamefont {Yang},
  \citenamefont {Kimura},\ and\ \citenamefont {Otani}}]{yang08}%
  \BibitemOpen
  \bibfield  {author} {\bibinfo {author} {\bibfnamefont {T.}~\bibnamefont
  {Yang}}, \bibinfo {author} {\bibfnamefont {T.}~\bibnamefont {Kimura}}, \ and\
  \bibinfo {author} {\bibfnamefont {Y.}~\bibnamefont {Otani}},\ }\bibfield
  {title} {\enquote {\bibinfo {title} {Giant spin-accumulation signal and pure
  spin-current-induced reversible magnetization switching},}\ }\href@noop {}
  {\bibfield  {journal} {\bibinfo  {journal} {Nat. Phys.}\ }\textbf {\bibinfo
  {volume} {4}},\ \bibinfo {pages} {851} (\bibinfo {year} {2008})}\BibitemShut
  {NoStop}%
\bibitem [{\citenamefont {Ando}\ \emph {et~al.}(2008)\citenamefont {Ando},
  \citenamefont {Takahashi}, \citenamefont {Harii}, \citenamefont {Sasage},
  \citenamefont {Ieda}, \citenamefont {Maekawa},\ and\ \citenamefont
  {Saitoh}}]{ando08}%
  \BibitemOpen
  \bibfield  {author} {\bibinfo {author} {\bibfnamefont {K.}~\bibnamefont
  {Ando}}, \bibinfo {author} {\bibfnamefont {S.}~\bibnamefont {Takahashi}},
  \bibinfo {author} {\bibfnamefont {K.}~\bibnamefont {Harii}}, \bibinfo
  {author} {\bibfnamefont {K.}~\bibnamefont {Sasage}}, \bibinfo {author}
  {\bibfnamefont {J.}~\bibnamefont {Ieda}}, \bibinfo {author} {\bibfnamefont
  {S.}~\bibnamefont {Maekawa}}, \ and\ \bibinfo {author} {\bibfnamefont
  {E.}~\bibnamefont {Saitoh}},\ }\bibfield  {title} {\enquote {\bibinfo {title}
  {Electric {Manipulation} of {Spin} {Relaxation} {Using} the {Spin} {Hall}
  {Effect}},}\ }\href@noop {} {\bibfield  {journal} {\bibinfo  {journal} {Phys.
  Rev. Lett.}\ }\textbf {\bibinfo {volume} {101}},\ \bibinfo {pages} {036601}
  (\bibinfo {year} {2008})}\BibitemShut {NoStop}%
\bibitem [{\citenamefont {Liu}\ \emph {et~al.}(2012{\natexlab{a}})\citenamefont
  {Liu}, \citenamefont {Pai}, \citenamefont {Li}, \citenamefont {Tseng},
  \citenamefont {Ralph},\ and\ \citenamefont {Buhrman}}]{liu12a}%
  \BibitemOpen
  \bibfield  {author} {\bibinfo {author} {\bibfnamefont {L.}~\bibnamefont
  {Liu}}, \bibinfo {author} {\bibfnamefont {C.-F.}\ \bibnamefont {Pai}},
  \bibinfo {author} {\bibfnamefont {Y.}~\bibnamefont {Li}}, \bibinfo {author}
  {\bibfnamefont {H.~W.}\ \bibnamefont {Tseng}}, \bibinfo {author}
  {\bibfnamefont {D.~C.}\ \bibnamefont {Ralph}}, \ and\ \bibinfo {author}
  {\bibfnamefont {R.~A.}\ \bibnamefont {Buhrman}},\ }\bibfield  {title}
  {\enquote {\bibinfo {title} {Spin-{Torque} {Switching} with the {Giant}
  {Spin} {Hall} {Effect} of {Tantalum}},}\ }\href@noop {} {\bibfield  {journal}
  {\bibinfo  {journal} {Science}\ }\textbf {\bibinfo {volume} {336}},\ \bibinfo
  {pages} {555} (\bibinfo {year} {2012}{\natexlab{a}})}\BibitemShut {NoStop}%
\bibitem [{\citenamefont {Liu}\ \emph {et~al.}(2012{\natexlab{b}})\citenamefont
  {Liu}, \citenamefont {Pai}, \citenamefont {Ralph},\ and\ \citenamefont
  {Buhrman}}]{liu12b}%
  \BibitemOpen
  \bibfield  {author} {\bibinfo {author} {\bibfnamefont {L.}~\bibnamefont
  {Liu}}, \bibinfo {author} {\bibfnamefont {C.-F.}\ \bibnamefont {Pai}},
  \bibinfo {author} {\bibfnamefont {D.~C.}\ \bibnamefont {Ralph}}, \ and\
  \bibinfo {author} {\bibfnamefont {R.~A.}\ \bibnamefont {Buhrman}},\
  }\bibfield  {title} {\enquote {\bibinfo {title} {Magnetization {Oscillations}
  {Drive} by the {Spin} {Hall} {Effect} in 3-{Terminal} {Magnetic} {Tunnel}
  {Junctions} {Devices}},}\ }\href@noop {} {\bibfield  {journal} {\bibinfo
  {journal} {Phys. Rev. Lett.}\ }\textbf {\bibinfo {volume} {109}},\ \bibinfo
  {pages} {186602} (\bibinfo {year} {2012}{\natexlab{b}})}\BibitemShut
  {NoStop}%
\bibitem [{\citenamefont {Pai}\ \emph {et~al.}(2012)\citenamefont {Pai},
  \citenamefont {Liu}, \citenamefont {Li}, \citenamefont {Tseng}, \citenamefont
  {Ralph},\ and\ \citenamefont {Buhrman}}]{pai12}%
  \BibitemOpen
  \bibfield  {author} {\bibinfo {author} {\bibfnamefont {C.-F.}\ \bibnamefont
  {Pai}}, \bibinfo {author} {\bibfnamefont {L.}~\bibnamefont {Liu}}, \bibinfo
  {author} {\bibfnamefont {Y.}~\bibnamefont {Li}}, \bibinfo {author}
  {\bibfnamefont {H.~W.}\ \bibnamefont {Tseng}}, \bibinfo {author}
  {\bibfnamefont {D.~C.}\ \bibnamefont {Ralph}}, \ and\ \bibinfo {author}
  {\bibfnamefont {R.~A.}\ \bibnamefont {Buhrman}},\ }\bibfield  {title}
  {\enquote {\bibinfo {title} {Spin transfer torque devices utilizing the giant
  spin {Hall} effect of tungsten},}\ }\href@noop {} {\bibfield  {journal}
  {\bibinfo  {journal} {Appl. Phys. Lett.}\ }\textbf {\bibinfo {volume}
  {101}},\ \bibinfo {pages} {122404} (\bibinfo {year} {2012})}\BibitemShut
  {NoStop}%
\bibitem [{\citenamefont {Yamanouchi}\ \emph {et~al.}(2013)\citenamefont
  {Yamanouchi}, \citenamefont {Chen}, \citenamefont {Kim}, \citenamefont
  {Hayashi}, \citenamefont {Sato}, \citenamefont {Fukami}, \citenamefont
  {Ikeda}, \citenamefont {Matsukura},\ and\ \citenamefont
  {Ohno}}]{yamanouchi13}%
  \BibitemOpen
  \bibfield  {author} {\bibinfo {author} {\bibfnamefont {M.}~\bibnamefont
  {Yamanouchi}}, \bibinfo {author} {\bibfnamefont {L.}~\bibnamefont {Chen}},
  \bibinfo {author} {\bibfnamefont {J.}~\bibnamefont {Kim}}, \bibinfo {author}
  {\bibfnamefont {M.}~\bibnamefont {Hayashi}}, \bibinfo {author} {\bibfnamefont
  {H.}~\bibnamefont {Sato}}, \bibinfo {author} {\bibfnamefont {S.}~\bibnamefont
  {Fukami}}, \bibinfo {author} {\bibfnamefont {S.}~\bibnamefont {Ikeda}},
  \bibinfo {author} {\bibfnamefont {F.}~\bibnamefont {Matsukura}}, \ and\
  \bibinfo {author} {\bibfnamefont {H.}~\bibnamefont {Ohno}},\ }\bibfield
  {title} {\enquote {\bibinfo {title} {Three terminal magnetic tunnel junction
  utilizing the spin {Hall} effect of iridium-doped copper},}\ }\href@noop {}
  {\bibfield  {journal} {\bibinfo  {journal} {Appl. Phys. Lett.}\ }\textbf
  {\bibinfo {volume} {102}},\ \bibinfo {pages} {212408} (\bibinfo {year}
  {2013})}\BibitemShut {NoStop}%
\bibitem [{\citenamefont {Fan}\ \emph {et~al.}(2013)\citenamefont {Fan},
  \citenamefont {Wu}, \citenamefont {Chen}, \citenamefont {Jerry},
  \citenamefont {Zhang},\ and\ \citenamefont {Xiao}}]{fan13}%
  \BibitemOpen
  \bibfield  {author} {\bibinfo {author} {\bibfnamefont {X.}~\bibnamefont
  {Fan}}, \bibinfo {author} {\bibfnamefont {J.}~\bibnamefont {Wu}}, \bibinfo
  {author} {\bibfnamefont {Y.}~\bibnamefont {Chen}}, \bibinfo {author}
  {\bibfnamefont {M.~J.}\ \bibnamefont {Jerry}}, \bibinfo {author}
  {\bibfnamefont {H.}~\bibnamefont {Zhang}}, \ and\ \bibinfo {author}
  {\bibfnamefont {J.~Q.}\ \bibnamefont {Xiao}},\ }\bibfield  {title} {\enquote
  {\bibinfo {title} {Observation of the nonlocal spin-orbital effective
  field},}\ }\href@noop {} {\bibfield  {journal} {\bibinfo  {journal} {Nat.
  Commun.}\ }\textbf {\bibinfo {volume} {4}},\ \bibinfo {pages} {1799}
  (\bibinfo {year} {2013})}\BibitemShut {NoStop}%
\bibitem [{\citenamefont {Cubukcu}\ \emph {et~al.}(2014)\citenamefont
  {Cubukcu}, \citenamefont {Boulle}, \citenamefont {Drouard}, \citenamefont
  {Garello}, \citenamefont {Avci}, \citenamefont {Miron}, \citenamefont
  {Langer}, \citenamefont {Ocker}, \citenamefont {Gambardella},\ and\
  \citenamefont {Gaudin}}]{cubukcu14}%
  \BibitemOpen
  \bibfield  {author} {\bibinfo {author} {\bibfnamefont {M.}~\bibnamefont
  {Cubukcu}}, \bibinfo {author} {\bibfnamefont {O.}~\bibnamefont {Boulle}},
  \bibinfo {author} {\bibfnamefont {M.}~\bibnamefont {Drouard}}, \bibinfo
  {author} {\bibfnamefont {K.}~\bibnamefont {Garello}}, \bibinfo {author}
  {\bibfnamefont {C.~O.}\ \bibnamefont {Avci}}, \bibinfo {author}
  {\bibfnamefont {I.~M.}\ \bibnamefont {Miron}}, \bibinfo {author}
  {\bibfnamefont {J.}~\bibnamefont {Langer}}, \bibinfo {author} {\bibfnamefont
  {B.}~\bibnamefont {Ocker}}, \bibinfo {author} {\bibfnamefont
  {P.}~\bibnamefont {Gambardella}}, \ and\ \bibinfo {author} {\bibfnamefont
  {G.}~\bibnamefont {Gaudin}},\ }\bibfield  {title} {\enquote {\bibinfo {title}
  {Spin-orbit torque magnetization switching of a three-terminal perpendicular
  magnetic tunnel junction},}\ }\href@noop {} {\bibfield  {journal} {\bibinfo
  {journal} {Appl. Phys. Lett.}\ }\textbf {\bibinfo {volume} {104}},\ \bibinfo
  {pages} {042406} (\bibinfo {year} {2014})}\BibitemShut {NoStop}%
\bibitem [{\citenamefont {Lee}\ \emph {et~al.}(2013)\citenamefont {Lee},
  \citenamefont {Lee}, \citenamefont {Min},\ and\ \citenamefont {Lee}}]{lee13}%
  \BibitemOpen
  \bibfield  {author} {\bibinfo {author} {\bibfnamefont {K.-S.}\ \bibnamefont
  {Lee}}, \bibinfo {author} {\bibfnamefont {S.-W.}\ \bibnamefont {Lee}},
  \bibinfo {author} {\bibfnamefont {B.-C.}\ \bibnamefont {Min}}, \ and\
  \bibinfo {author} {\bibfnamefont {K.-J.}\ \bibnamefont {Lee}},\ }\bibfield
  {title} {\enquote {\bibinfo {title} {Threshold current for switching of a
  perpendicular magnetic layer induced by spin {Hall} effect},}\ }\href@noop {}
  {\bibfield  {journal} {\bibinfo  {journal} {Appl. Phys. Lett.}\ }\textbf
  {\bibinfo {volume} {102}},\ \bibinfo {pages} {112410} (\bibinfo {year}
  {2013})}\BibitemShut {NoStop}%
\bibitem [{\citenamefont {Lee}\ \emph {et~al.}(2014)\citenamefont {Lee},
  \citenamefont {Lee}, \citenamefont {Min},\ and\ \citenamefont {Lee}}]{lee14}%
  \BibitemOpen
  \bibfield  {author} {\bibinfo {author} {\bibfnamefont {K.-S.}\ \bibnamefont
  {Lee}}, \bibinfo {author} {\bibfnamefont {S.-W.}\ \bibnamefont {Lee}},
  \bibinfo {author} {\bibfnamefont {B.-C.}\ \bibnamefont {Min}}, \ and\
  \bibinfo {author} {\bibfnamefont {K.-J.}\ \bibnamefont {Lee}},\ }\bibfield
  {title} {\enquote {\bibinfo {title} {Thermally activated switching of
  perpendicular magnet by spin-orbit spin torque},}\ }\href@noop {} {\bibfield
  {journal} {\bibinfo  {journal} {Appl. Phys. Lett.}\ }\textbf {\bibinfo
  {volume} {104}},\ \bibinfo {pages} {072413} (\bibinfo {year}
  {2014})}\BibitemShut {NoStop}%
\bibitem [{\citenamefont {Sun}(2000)}]{sun00}%
  \BibitemOpen
  \bibfield  {author} {\bibinfo {author} {\bibfnamefont {J.~Z.}\ \bibnamefont
  {Sun}},\ }\bibfield  {title} {\enquote {\bibinfo {title} {Spin-current
  interaction with a monodomain magnetic body: {A} model study},}\ }\href@noop
  {} {\bibfield  {journal} {\bibinfo  {journal} {Phys. Rev. B}\ }\textbf
  {\bibinfo {volume} {62}},\ \bibinfo {pages} {570} (\bibinfo {year}
  {2000})}\BibitemShut {NoStop}%
\bibitem [{\citenamefont {Ikeda}\ \emph {et~al.}(2010)\citenamefont {Ikeda},
  \citenamefont {Miura}, \citenamefont {Yamamoto}, \citenamefont {Mizunuma},
  \citenamefont {Gan}, \citenamefont {Endo}, \citenamefont {Kanai},
  \citenamefont {Hayakawa}, \citenamefont {Matsukura},\ and\ \citenamefont
  {Ohno}}]{ikeda10}%
  \BibitemOpen
  \bibfield  {author} {\bibinfo {author} {\bibfnamefont {S.}~\bibnamefont
  {Ikeda}}, \bibinfo {author} {\bibfnamefont {K.}~\bibnamefont {Miura}},
  \bibinfo {author} {\bibfnamefont {H}~\bibnamefont {Yamamoto}}, \bibinfo
  {author} {\bibfnamefont {K.}~\bibnamefont {Mizunuma}}, \bibinfo {author}
  {\bibfnamefont {H.~D.}\ \bibnamefont {Gan}}, \bibinfo {author} {\bibfnamefont
  {M.}~\bibnamefont {Endo}}, \bibinfo {author} {\bibfnamefont {S.}~\bibnamefont
  {Kanai}}, \bibinfo {author} {\bibfnamefont {J.}~\bibnamefont {Hayakawa}},
  \bibinfo {author} {\bibfnamefont {F.}~\bibnamefont {Matsukura}}, \ and\
  \bibinfo {author} {\bibfnamefont {H.}~\bibnamefont {Ohno}},\ }\bibfield
  {title} {\enquote {\bibinfo {title} {A perpendicular-anisotropy
  {C}o{F}e{B}-{M}g{O} magnetic tunnel junction},}\ }\href@noop {} {\bibfield
  {journal} {\bibinfo  {journal} {Nat. Mater.}\ }\textbf {\bibinfo {volume}
  {9}},\ \bibinfo {pages} {721} (\bibinfo {year} {2010})}\BibitemShut {NoStop}%
\bibitem [{\citenamefont {Iihama}\ \emph {et~al.}(2012)\citenamefont {Iihama},
  \citenamefont {Ma}, \citenamefont {Kubota}, \citenamefont {Mizukami},
  \citenamefont {Ando},\ and\ \citenamefont {Miyazaki}}]{iihama12}%
  \BibitemOpen
  \bibfield  {author} {\bibinfo {author} {\bibfnamefont {S.}~\bibnamefont
  {Iihama}}, \bibinfo {author} {\bibfnamefont {Q.}~\bibnamefont {Ma}}, \bibinfo
  {author} {\bibfnamefont {T.}~\bibnamefont {Kubota}}, \bibinfo {author}
  {\bibfnamefont {S.}~\bibnamefont {Mizukami}}, \bibinfo {author}
  {\bibfnamefont {Y.}~\bibnamefont {Ando}}, \ and\ \bibinfo {author}
  {\bibfnamefont {T.}~\bibnamefont {Miyazaki}},\ }\bibfield  {title} {\enquote
  {\bibinfo {title} {Damping of {Magnetization} {Precession} in
  {Perpendicularly} {Magnetized} {Co}{Fe}{B} {Alloy} {Thin} {Films}},}\
  }\href@noop {} {\bibfield  {journal} {\bibinfo  {journal} {Appl. Phys.
  Express}\ }\textbf {\bibinfo {volume} {5}},\ \bibinfo {pages} {083001}
  (\bibinfo {year} {2012})}\BibitemShut {NoStop}%
\bibitem [{\citenamefont {Zhang}\ \emph {et~al.}(2002)\citenamefont {Zhang},
  \citenamefont {Levy},\ and\ \citenamefont {Fert}}]{zhang02}%
  \BibitemOpen
  \bibfield  {author} {\bibinfo {author} {\bibfnamefont {S.}~\bibnamefont
  {Zhang}}, \bibinfo {author} {\bibfnamefont {P.~M.}\ \bibnamefont {Levy}}, \
  and\ \bibinfo {author} {\bibfnamefont {A.}~\bibnamefont {Fert}},\ }\bibfield
  {title} {\enquote {\bibinfo {title} {Mechanisms of {Spin}-{Polarized}
  {Current}-{Driven} {Magnetization} {Switching}},}\ }\href@noop {} {\bibfield
  {journal} {\bibinfo  {journal} {Phys. Rev. Lett.}\ }\textbf {\bibinfo
  {volume} {88}},\ \bibinfo {pages} {236601} (\bibinfo {year}
  {2002})}\BibitemShut {NoStop}%
\bibitem [{\citenamefont {Stiles}\ and\ \citenamefont
  {Zangwill}(2002)}]{stiles02}%
  \BibitemOpen
  \bibfield  {author} {\bibinfo {author} {\bibfnamefont {M.~D.}\ \bibnamefont
  {Stiles}}\ and\ \bibinfo {author} {\bibfnamefont {A.}~\bibnamefont
  {Zangwill}},\ }\bibfield  {title} {\enquote {\bibinfo {title} {Anatomy of
  spin-transfer torque},}\ }\href@noop {} {\bibfield  {journal} {\bibinfo
  {journal} {Phys. Rev. B}\ }\textbf {\bibinfo {volume} {66}},\ \bibinfo
  {pages} {014407} (\bibinfo {year} {2002})}\BibitemShut {NoStop}%
\bibitem [{\citenamefont {Zwierzycki}\ \emph {et~al.}(2005)\citenamefont
  {Zwierzycki}, \citenamefont {Tserkovnyak}, \citenamefont {Kelly},
  \citenamefont {Brataas},\ and\ \citenamefont {Bauer}}]{zwierzycki05}%
  \BibitemOpen
  \bibfield  {author} {\bibinfo {author} {\bibfnamefont {M.}~\bibnamefont
  {Zwierzycki}}, \bibinfo {author} {\bibfnamefont {Y.}~\bibnamefont
  {Tserkovnyak}}, \bibinfo {author} {\bibfnamefont {P.~J.}\ \bibnamefont
  {Kelly}}, \bibinfo {author} {\bibfnamefont {A.}~\bibnamefont {Brataas}}, \
  and\ \bibinfo {author} {\bibfnamefont {G.~E.~W.}\ \bibnamefont {Bauer}},\
  }\bibfield  {title} {\enquote {\bibinfo {title} {First-principles study of
  magnetization relaxation enhancement and spin transfer in thin magnetic
  films},}\ }\href@noop {} {\bibfield  {journal} {\bibinfo  {journal} {Phys.
  Rev. B}\ }\textbf {\bibinfo {volume} {71}},\ \bibinfo {pages} {064420}
  (\bibinfo {year} {2005})}\BibitemShut {NoStop}%
\bibitem [{\citenamefont {Brataas}\ \emph {et~al.}(2006)\citenamefont
  {Brataas}, \citenamefont {Bauer},\ and\ \citenamefont {Kelly}}]{brataas06}%
  \BibitemOpen
  \bibfield  {author} {\bibinfo {author} {\bibfnamefont {A.}~\bibnamefont
  {Brataas}}, \bibinfo {author} {\bibfnamefont {G.~E.~W.}\ \bibnamefont
  {Bauer}}, \ and\ \bibinfo {author} {\bibfnamefont {P.~J.}\ \bibnamefont
  {Kelly}},\ }\bibfield  {title} {\enquote {\bibinfo {title} {Non-collinear
  magnetoelectronics},}\ }\href@noop {} {\bibfield  {journal} {\bibinfo
  {journal} {Phys. Rep.}\ }\textbf {\bibinfo {volume} {427}},\ \bibinfo {pages}
  {157} (\bibinfo {year} {2006})}\BibitemShut {NoStop}%
\bibitem [{\citenamefont {Theodonis}\ \emph {et~al.}(2006)\citenamefont
  {Theodonis}, \citenamefont {Kioussis}, \citenamefont {Kalitsov},
  \citenamefont {Chshiev},\ and\ \citenamefont {Butler}}]{theodonis06}%
  \BibitemOpen
  \bibfield  {author} {\bibinfo {author} {\bibfnamefont {I.}~\bibnamefont
  {Theodonis}}, \bibinfo {author} {\bibfnamefont {N.}~\bibnamefont {Kioussis}},
  \bibinfo {author} {\bibfnamefont {A.}~\bibnamefont {Kalitsov}}, \bibinfo
  {author} {\bibfnamefont {M.}~\bibnamefont {Chshiev}}, \ and\ \bibinfo
  {author} {\bibfnamefont {W.~H.}\ \bibnamefont {Butler}},\ }\bibfield  {title}
  {\enquote {\bibinfo {title} {Anomalous {B}ias {D}ependence of {S}pin {T}orque
  in {M}agnetic {T}unnel {J}unctions},}\ }\href@noop {} {\bibfield  {journal}
  {\bibinfo  {journal} {Phys. Rev. Lett.}\ }\textbf {\bibinfo {volume} {97}},\
  \bibinfo {pages} {237205} (\bibinfo {year} {2006})}\BibitemShut {NoStop}%
\bibitem [{\citenamefont {Taniguchi}\ \emph {et~al.}(2008)\citenamefont
  {Taniguchi}, \citenamefont {Yakata}, \citenamefont {Imamura},\ and\
  \citenamefont {Ando}}]{taniguchi08}%
  \BibitemOpen
  \bibfield  {author} {\bibinfo {author} {\bibfnamefont {T.}~\bibnamefont
  {Taniguchi}}, \bibinfo {author} {\bibfnamefont {S.}~\bibnamefont {Yakata}},
  \bibinfo {author} {\bibfnamefont {H.}~\bibnamefont {Imamura}}, \ and\
  \bibinfo {author} {\bibfnamefont {Y.}~\bibnamefont {Ando}},\ }\bibfield
  {title} {\enquote {\bibinfo {title} {Determination of {Penetration} {Depth}
  of {Transverse} {Spin} {Current} in {Ferromagnetic} {Metals} by {Spin}
  {Pumping}},}\ }\href@noop {} {\bibfield  {journal} {\bibinfo  {journal}
  {Appl. Phys. Express}\ }\textbf {\bibinfo {volume} {1}},\ \bibinfo {pages}
  {031302} (\bibinfo {year} {2008})}\BibitemShut {NoStop}%
\bibitem [{\citenamefont {Ghosh}\ \emph {et~al.}(2012)\citenamefont {Ghosh},
  \citenamefont {Auffret}, \citenamefont {Ebels},\ and\ \citenamefont
  {Bailey}}]{ghosh12}%
  \BibitemOpen
  \bibfield  {author} {\bibinfo {author} {\bibfnamefont {A.}~\bibnamefont
  {Ghosh}}, \bibinfo {author} {\bibfnamefont {S.}~\bibnamefont {Auffret}},
  \bibinfo {author} {\bibfnamefont {U.}~\bibnamefont {Ebels}}, \ and\ \bibinfo
  {author} {\bibfnamefont {W.~E.}\ \bibnamefont {Bailey}},\ }\bibfield  {title}
  {\enquote {\bibinfo {title} {Penetration {Depth} of {Transverse} {Spin}
  {Current} in {Ultrahin} {Ferromagnets}},}\ }\href@noop {} {\bibfield
  {journal} {\bibinfo  {journal} {Phys. Rev. Lett.}\ }\textbf {\bibinfo
  {volume} {109}},\ \bibinfo {pages} {127202} (\bibinfo {year}
  {2012})}\BibitemShut {NoStop}%
\bibitem [{\citenamefont {Tulapurkar}\ \emph {et~al.}(2005)\citenamefont
  {Tulapurkar}, \citenamefont {Suzuki}, \citenamefont {Fukushima},
  \citenamefont {Kubota}, \citenamefont {Maehara}, \citenamefont {Tsunekawa},
  \citenamefont {Djayaprawira}, \citenamefont {Watanabe},\ and\ \citenamefont
  {Yuasa}}]{tulapurkar05}%
  \BibitemOpen
  \bibfield  {author} {\bibinfo {author} {\bibfnamefont {A.~A.}\ \bibnamefont
  {Tulapurkar}}, \bibinfo {author} {\bibfnamefont {Y.}~\bibnamefont {Suzuki}},
  \bibinfo {author} {\bibfnamefont {A.}~\bibnamefont {Fukushima}}, \bibinfo
  {author} {\bibfnamefont {H.}~\bibnamefont {Kubota}}, \bibinfo {author}
  {\bibfnamefont {H.}~\bibnamefont {Maehara}}, \bibinfo {author} {\bibfnamefont
  {K.}~\bibnamefont {Tsunekawa}}, \bibinfo {author} {\bibfnamefont {D.~D.}\
  \bibnamefont {Djayaprawira}}, \bibinfo {author} {\bibfnamefont
  {N.}~\bibnamefont {Watanabe}}, \ and\ \bibinfo {author} {\bibfnamefont
  {S.}~\bibnamefont {Yuasa}},\ }\bibfield  {title} {\enquote {\bibinfo {title}
  {Spin-torque diode effect in magnetic tunnel junctions},}\ }\href@noop {}
  {\bibfield  {journal} {\bibinfo  {journal} {Nature}\ }\textbf {\bibinfo
  {volume} {438}},\ \bibinfo {pages} {339} (\bibinfo {year}
  {2005})}\BibitemShut {NoStop}%
\bibitem [{\citenamefont {Kubota}\ \emph {et~al.}(2008)\citenamefont {Kubota},
  \citenamefont {Fukushima}, \citenamefont {Yakushiji}, \citenamefont
  {Nagahama}, \citenamefont {Yuasa}, \citenamefont {Ando}, \citenamefont
  {Maehara}, \citenamefont {Nagamine}, \citenamefont {Tsunekawa}, \citenamefont
  {Djayaprawira}, \citenamefont {Watanabe},\ and\ \citenamefont
  {Suzuki}}]{kubota08}%
  \BibitemOpen
  \bibfield  {author} {\bibinfo {author} {\bibfnamefont {H.}~\bibnamefont
  {Kubota}}, \bibinfo {author} {\bibfnamefont {A.}~\bibnamefont {Fukushima}},
  \bibinfo {author} {\bibfnamefont {K.}~\bibnamefont {Yakushiji}}, \bibinfo
  {author} {\bibfnamefont {T.}~\bibnamefont {Nagahama}}, \bibinfo {author}
  {\bibfnamefont {S.}~\bibnamefont {Yuasa}}, \bibinfo {author} {\bibfnamefont
  {K.}~\bibnamefont {Ando}}, \bibinfo {author} {\bibfnamefont {H.}~\bibnamefont
  {Maehara}}, \bibinfo {author} {\bibfnamefont {Y.}~\bibnamefont {Nagamine}},
  \bibinfo {author} {\bibfnamefont {K.}~\bibnamefont {Tsunekawa}}, \bibinfo
  {author} {\bibfnamefont {D.~D.}\ \bibnamefont {Djayaprawira}}, \bibinfo
  {author} {\bibfnamefont {N.}~\bibnamefont {Watanabe}}, \ and\ \bibinfo
  {author} {\bibfnamefont {Y.}~\bibnamefont {Suzuki}},\ }\bibfield  {title}
  {\enquote {\bibinfo {title} {Quantitative measurement of voltage dependence
  of spin-transfer-torque in {Mg}{O}-based magnetic tunnel junctions},}\
  }\href@noop {} {\bibfield  {journal} {\bibinfo  {journal} {Nat. Phys.}\
  }\textbf {\bibinfo {volume} {4}},\ \bibinfo {pages} {37} (\bibinfo {year}
  {2008})}\BibitemShut {NoStop}%
\bibitem [{\citenamefont {Sankey}\ \emph {et~al.}(2008)\citenamefont {Sankey},
  \citenamefont {Cui}, \citenamefont {Sun}, \citenamefont {Slonczewski},
  \citenamefont {Buhrman},\ and\ \citenamefont {Ralph}}]{sankey08}%
  \BibitemOpen
  \bibfield  {author} {\bibinfo {author} {\bibfnamefont {J.~C.}\ \bibnamefont
  {Sankey}}, \bibinfo {author} {\bibfnamefont {Y.-T.}\ \bibnamefont {Cui}},
  \bibinfo {author} {\bibfnamefont {J.~Z.}\ \bibnamefont {Sun}}, \bibinfo
  {author} {\bibfnamefont {J.~C.}\ \bibnamefont {Slonczewski}}, \bibinfo
  {author} {\bibfnamefont {R.~A.}\ \bibnamefont {Buhrman}}, \ and\ \bibinfo
  {author} {\bibfnamefont {D.~C.}\ \bibnamefont {Ralph}},\ }\bibfield  {title}
  {\enquote {\bibinfo {title} {Measurement of the spin-transfer-torque vector
  in magnetic tunnel junctions},}\ }\href@noop {} {\bibfield  {journal}
  {\bibinfo  {journal} {Nat. Phys.}\ }\textbf {\bibinfo {volume} {4}},\
  \bibinfo {pages} {67} (\bibinfo {year} {2008})}\BibitemShut {NoStop}%
\bibitem [{\citenamefont {Oh}\ \emph {et~al.}(2009)\citenamefont {Oh},
  \citenamefont {Park}, \citenamefont {Manchon}, \citenamefont {Chshiev},
  \citenamefont {Han}, \citenamefont {Lee}, \citenamefont {Lee}, \citenamefont
  {Nam}, \citenamefont {Jo}, \citenamefont {Kong}, \citenamefont {Dieny},\ and\
  \citenamefont {Lee}}]{oh09}%
  \BibitemOpen
  \bibfield  {author} {\bibinfo {author} {\bibfnamefont {S.-C.}\ \bibnamefont
  {Oh}}, \bibinfo {author} {\bibfnamefont {S.-Y.}\ \bibnamefont {Park}},
  \bibinfo {author} {\bibfnamefont {A.}~\bibnamefont {Manchon}}, \bibinfo
  {author} {\bibfnamefont {M.}~\bibnamefont {Chshiev}}, \bibinfo {author}
  {\bibfnamefont {J.-H.}\ \bibnamefont {Han}}, \bibinfo {author} {\bibfnamefont
  {H.-W.}\ \bibnamefont {Lee}}, \bibinfo {author} {\bibfnamefont {J.-E.}\
  \bibnamefont {Lee}}, \bibinfo {author} {\bibfnamefont {K.-T.}\ \bibnamefont
  {Nam}}, \bibinfo {author} {\bibfnamefont {Y.}~\bibnamefont {Jo}}, \bibinfo
  {author} {\bibfnamefont {Y.-C.}\ \bibnamefont {Kong}}, \bibinfo {author}
  {\bibfnamefont {B.}~\bibnamefont {Dieny}}, \ and\ \bibinfo {author}
  {\bibfnamefont {K.-J.}\ \bibnamefont {Lee}},\ }\bibfield  {title} {\enquote
  {\bibinfo {title} {Bias-voltage dependence of perpendicular spin-transfer
  torque in asymmetric {Mg}{O}-based magnetic tunnel junctions},}\ }\href@noop
  {} {\bibfield  {journal} {\bibinfo  {journal} {Nat. Phys.}\ }\textbf
  {\bibinfo {volume} {5}},\ \bibinfo {pages} {898} (\bibinfo {year}
  {2009})}\BibitemShut {NoStop}%
\bibitem [{\citenamefont {Grollier}\ \emph {et~al.}(2003)\citenamefont
  {Grollier}, \citenamefont {Cros}, \citenamefont {Jaffr\'es}, \citenamefont
  {Hamzic}, \citenamefont {George}, \citenamefont {Faini}, \citenamefont
  {Youssef}, \citenamefont {LeGall},\ and\ \citenamefont {Fert}}]{grollier03}%
  \BibitemOpen
  \bibfield  {author} {\bibinfo {author} {\bibfnamefont {J.}~\bibnamefont
  {Grollier}}, \bibinfo {author} {\bibfnamefont {V.}~\bibnamefont {Cros}},
  \bibinfo {author} {\bibfnamefont {H.}~\bibnamefont {Jaffr\'es}}, \bibinfo
  {author} {\bibfnamefont {A.}~\bibnamefont {Hamzic}}, \bibinfo {author}
  {\bibfnamefont {J.~M.}\ \bibnamefont {George}}, \bibinfo {author}
  {\bibfnamefont {G.}~\bibnamefont {Faini}}, \bibinfo {author} {\bibfnamefont
  {J.~Ben}\ \bibnamefont {Youssef}}, \bibinfo {author} {\bibfnamefont
  {H.}~\bibnamefont {LeGall}}, \ and\ \bibinfo {author} {\bibfnamefont
  {A.}~\bibnamefont {Fert}},\ }\bibfield  {title} {\enquote {\bibinfo {title}
  {Field dependence of magnetization reversal by spin transfer},}\ }\href@noop
  {} {\bibfield  {journal} {\bibinfo  {journal} {Phys. Rev. B}\ }\textbf
  {\bibinfo {volume} {67}},\ \bibinfo {pages} {174402} (\bibinfo {year}
  {2003})}\BibitemShut {NoStop}%
\bibitem [{\citenamefont {Morise}\ and\ \citenamefont
  {Nakamura}(2005)}]{morise05}%
  \BibitemOpen
  \bibfield  {author} {\bibinfo {author} {\bibfnamefont {H.}~\bibnamefont
  {Morise}}\ and\ \bibinfo {author} {\bibfnamefont {S.}~\bibnamefont
  {Nakamura}},\ }\bibfield  {title} {\enquote {\bibinfo {title} {Stable
  magnetization states under a spin-polarized current and a magnetic field},}\
  }\href@noop {} {\bibfield  {journal} {\bibinfo  {journal} {Phys. Rev. B}\
  }\textbf {\bibinfo {volume} {71}},\ \bibinfo {pages} {014439} (\bibinfo
  {year} {2005})}\BibitemShut {NoStop}%
\bibitem [{\citenamefont {Gusakova}\ \emph {et~al.}(2009)\citenamefont
  {Gusakova}, \citenamefont {Houssameddine}, \citenamefont {Ebels},
  \citenamefont {Dieny}, \citenamefont {Buda-Prejbeanu}, \citenamefont
  {Cyrille},\ and\ \citenamefont {Dela\"et}}]{gusakova09}%
  \BibitemOpen
  \bibfield  {author} {\bibinfo {author} {\bibfnamefont {D.}~\bibnamefont
  {Gusakova}}, \bibinfo {author} {\bibfnamefont {D.}~\bibnamefont
  {Houssameddine}}, \bibinfo {author} {\bibfnamefont {U.}~\bibnamefont
  {Ebels}}, \bibinfo {author} {\bibfnamefont {B.}~\bibnamefont {Dieny}},
  \bibinfo {author} {\bibfnamefont {L.}~\bibnamefont {Buda-Prejbeanu}},
  \bibinfo {author} {\bibfnamefont {M.~C.}\ \bibnamefont {Cyrille}}, \ and\
  \bibinfo {author} {\bibfnamefont {B.}~\bibnamefont {Dela\"et}},\ }\bibfield
  {title} {\enquote {\bibinfo {title} {Spin-polarized current-induced
  excitations in a coupled magnetic layer system},}\ }\href@noop {} {\bibfield
  {journal} {\bibinfo  {journal} {Phys. Rev. B}\ }\textbf {\bibinfo {volume}
  {79}},\ \bibinfo {pages} {104406} (\bibinfo {year} {2009})}\BibitemShut
  {NoStop}%
\bibitem [{\citenamefont {Taniguchi}\ \emph {et~al.}(2014)\citenamefont
  {Taniguchi}, \citenamefont {Tsunegi}, \citenamefont {Kubota},\ and\
  \citenamefont {Imamura}}]{taniguchi14APL}%
  \BibitemOpen
  \bibfield  {author} {\bibinfo {author} {\bibfnamefont {T.}~\bibnamefont
  {Taniguchi}}, \bibinfo {author} {\bibfnamefont {S.}~\bibnamefont {Tsunegi}},
  \bibinfo {author} {\bibfnamefont {H.}~\bibnamefont {Kubota}}, \ and\ \bibinfo
  {author} {\bibfnamefont {H.}~\bibnamefont {Imamura}},\ }\bibfield  {title}
  {\enquote {\bibinfo {title} {Self-oscillation in spin torque oscillator
  stabilized by field-like torque},}\ }\href@noop {} {\bibfield  {journal}
  {\bibinfo  {journal} {Appl. Phys. Lett.}\ }\textbf {\bibinfo {volume}
  {104}},\ \bibinfo {pages} {152411} (\bibinfo {year} {2014})}\BibitemShut
  {NoStop}%
\bibitem [{\citenamefont {Taniguchi}\ \emph {et~al.}(2015)\citenamefont
  {Taniguchi}, \citenamefont {Tsunegi}, \citenamefont {Kubota},\ and\
  \citenamefont {Imamura}}]{taniguchi15JAP}%
  \BibitemOpen
  \bibfield  {author} {\bibinfo {author} {\bibfnamefont {T.}~\bibnamefont
  {Taniguchi}}, \bibinfo {author} {\bibfnamefont {S.}~\bibnamefont {Tsunegi}},
  \bibinfo {author} {\bibfnamefont {H.}~\bibnamefont {Kubota}}, \ and\ \bibinfo
  {author} {\bibfnamefont {H.}~\bibnamefont {Imamura}},\ }\bibfield  {title}
  {\enquote {\bibinfo {title} {Large amplitude oscillation of magnetization in
  spin-torque oscillator stabilized by field-like torque},}\ }\href@noop {}
  {\bibfield  {journal} {\bibinfo  {journal} {J. Appl. Phys.}\ }\textbf
  {\bibinfo {volume} {117}},\ \bibinfo {pages} {17C504} (\bibinfo {year}
  {2015})}\BibitemShut {NoStop}%
\bibitem [{\citenamefont {Kim}\ \emph {et~al.}(2013)\citenamefont {Kim},
  \citenamefont {Sinha}, \citenamefont {Hayashi}, \citenamefont {Yamanouchi},
  \citenamefont {Fukami}, \citenamefont {Suzuki}, \citenamefont {Mitani},\ and\
  \citenamefont {Ohno}}]{kim13}%
  \BibitemOpen
  \bibfield  {author} {\bibinfo {author} {\bibfnamefont {J.}~\bibnamefont
  {Kim}}, \bibinfo {author} {\bibfnamefont {J.}~\bibnamefont {Sinha}}, \bibinfo
  {author} {\bibfnamefont {M.}~\bibnamefont {Hayashi}}, \bibinfo {author}
  {\bibfnamefont {M.}~\bibnamefont {Yamanouchi}}, \bibinfo {author}
  {\bibfnamefont {S.}~\bibnamefont {Fukami}}, \bibinfo {author} {\bibfnamefont
  {T.}~\bibnamefont {Suzuki}}, \bibinfo {author} {\bibfnamefont
  {S.}~\bibnamefont {Mitani}}, \ and\ \bibinfo {author} {\bibfnamefont
  {H.}~\bibnamefont {Ohno}},\ }\bibfield  {title} {\enquote {\bibinfo {title}
  {Layer thickness dependence of the current-induced effective field vector in
  {Ta}$|${Co}{Fe}{B}$|${Mg}{O}},}\ }\href@noop {} {\bibfield  {journal}
  {\bibinfo  {journal} {Nat. Mater.}\ }\textbf {\bibinfo {volume} {12}},\
  \bibinfo {pages} {240} (\bibinfo {year} {2013})}\BibitemShut {NoStop}%
\bibitem [{\citenamefont {Garello}\ \emph {et~al.}(2013)\citenamefont
  {Garello}, \citenamefont {Miron}, \citenamefont {Avci}, \citenamefont
  {Freimuth}, \citenamefont {Mokrousov}, \citenamefont {Bl\"ugel},
  \citenamefont {Auffret}, \citenamefont {Boulle}, \citenamefont {Gaudin},\
  and\ \citenamefont {Gambardella}}]{garello13}%
  \BibitemOpen
  \bibfield  {author} {\bibinfo {author} {\bibfnamefont {K.}~\bibnamefont
  {Garello}}, \bibinfo {author} {\bibfnamefont {I.~M.}\ \bibnamefont {Miron}},
  \bibinfo {author} {\bibfnamefont {C.~O.}\ \bibnamefont {Avci}}, \bibinfo
  {author} {\bibfnamefont {F.}~\bibnamefont {Freimuth}}, \bibinfo {author}
  {\bibfnamefont {Y.}~\bibnamefont {Mokrousov}}, \bibinfo {author}
  {\bibfnamefont {S.}~\bibnamefont {Bl\"ugel}}, \bibinfo {author}
  {\bibfnamefont {S.}~\bibnamefont {Auffret}}, \bibinfo {author} {\bibfnamefont
  {O.}~\bibnamefont {Boulle}}, \bibinfo {author} {\bibfnamefont
  {G.}~\bibnamefont {Gaudin}}, \ and\ \bibinfo {author} {\bibfnamefont
  {P.}~\bibnamefont {Gambardella}},\ }\bibfield  {title} {\enquote {\bibinfo
  {title} {Symmetry and magnitude of spin-orbit torques in ferromagnetic
  heterostructures},}\ }\href@noop {} {\bibfield  {journal} {\bibinfo
  {journal} {Nat. Nanotech.}\ }\textbf {\bibinfo {volume} {8}},\ \bibinfo
  {pages} {587} (\bibinfo {year} {2013})}\BibitemShut {NoStop}%
\bibitem [{\citenamefont {Qiu}\ \emph {et~al.}(2014)\citenamefont {Qiu},
  \citenamefont {Deorani}, \citenamefont {Narayanapillai}, \citenamefont {Lee},
  \citenamefont {Lee}, \citenamefont {Lee},\ and\ \citenamefont
  {Yang}}]{qiu14}%
  \BibitemOpen
  \bibfield  {author} {\bibinfo {author} {\bibfnamefont {X.}~\bibnamefont
  {Qiu}}, \bibinfo {author} {\bibfnamefont {P.}~\bibnamefont {Deorani}},
  \bibinfo {author} {\bibfnamefont {K.}~\bibnamefont {Narayanapillai}},
  \bibinfo {author} {\bibfnamefont {K.-S.}\ \bibnamefont {Lee}}, \bibinfo
  {author} {\bibfnamefont {K.-J.}\ \bibnamefont {Lee}}, \bibinfo {author}
  {\bibfnamefont {H.-W.}\ \bibnamefont {Lee}}, \ and\ \bibinfo {author}
  {\bibfnamefont {H.}~\bibnamefont {Yang}},\ }\bibfield  {title} {\enquote
  {\bibinfo {title} {Angular and temperature dependence of current induced
  spin-orbit effective fields in {Ta}/{Co}{Fe}{B}/{Mg}{O} nanowires},}\
  }\href@noop {} {\bibfield  {journal} {\bibinfo  {journal} {Sci. Rep.}\
  }\textbf {\bibinfo {volume} {4}},\ \bibinfo {pages} {4491} (\bibinfo {year}
  {2014})}\BibitemShut {NoStop}%
\bibitem [{\citenamefont {Pai}\ \emph {et~al.}(2014)\citenamefont {Pai},
  \citenamefont {Nguyen}, \citenamefont {Belvin}, \citenamefont {Viela-Leao},
  \citenamefont {Ralph},\ and\ \citenamefont {Buhrman}}]{pai14}%
  \BibitemOpen
  \bibfield  {author} {\bibinfo {author} {\bibfnamefont {C.-F.}\ \bibnamefont
  {Pai}}, \bibinfo {author} {\bibfnamefont {M.-H.}\ \bibnamefont {Nguyen}},
  \bibinfo {author} {\bibfnamefont {C.}~\bibnamefont {Belvin}}, \bibinfo
  {author} {\bibfnamefont {L.~H.}\ \bibnamefont {Viela-Leao}}, \bibinfo
  {author} {\bibfnamefont {D.~C.}\ \bibnamefont {Ralph}}, \ and\ \bibinfo
  {author} {\bibfnamefont {R.~A.}\ \bibnamefont {Buhrman}},\ }\bibfield
  {title} {\enquote {\bibinfo {title} {Enhancement of perpendicular magnetic
  anisotropy and tranmission of spin-{Hall}-effect-induced spin currents by a
  {Hf} spacer layer in {W}/{Hf}/{Co}{Fe}{B}/{Mg}{O} layer structures},}\
  }\href@noop {} {\bibfield  {journal} {\bibinfo  {journal} {Appl. Phys.
  Lett.}\ }\textbf {\bibinfo {volume} {104}},\ \bibinfo {pages} {082407}
  (\bibinfo {year} {2014})}\BibitemShut {NoStop}%
\bibitem [{\citenamefont {Kim}\ \emph {et~al.}(2014)\citenamefont {Kim},
  \citenamefont {Sinha}, \citenamefont {Mitani}, \citenamefont {Hayashi},
  \citenamefont {Takahashi}, \citenamefont {Maekawa}, \citenamefont
  {Yamanouchi},\ and\ \citenamefont {Ohno}}]{kim14}%
  \BibitemOpen
  \bibfield  {author} {\bibinfo {author} {\bibfnamefont {J.}~\bibnamefont
  {Kim}}, \bibinfo {author} {\bibfnamefont {J.}~\bibnamefont {Sinha}}, \bibinfo
  {author} {\bibfnamefont {S.}~\bibnamefont {Mitani}}, \bibinfo {author}
  {\bibfnamefont {M.}~\bibnamefont {Hayashi}}, \bibinfo {author} {\bibfnamefont
  {S.}~\bibnamefont {Takahashi}}, \bibinfo {author} {\bibfnamefont
  {S.}~\bibnamefont {Maekawa}}, \bibinfo {author} {\bibfnamefont
  {M.}~\bibnamefont {Yamanouchi}}, \ and\ \bibinfo {author} {\bibfnamefont
  {H.}~\bibnamefont {Ohno}},\ }\bibfield  {title} {\enquote {\bibinfo {title}
  {Anomalous temperature dependence of current-induced torques in
  {Co}{Fe}{B}/{Mg}{O} heterostructures with {Ta}-based underlayers},}\
  }\href@noop {} {\bibfield  {journal} {\bibinfo  {journal} {Phys. Rev. B}\
  }\textbf {\bibinfo {volume} {89}},\ \bibinfo {pages} {174424} (\bibinfo
  {year} {2014})}\BibitemShut {NoStop}%
\bibitem [{\citenamefont {Torrejon}\ \emph {et~al.}(2014)\citenamefont
  {Torrejon}, \citenamefont {Kim}, \citenamefont {Sinha}, \citenamefont
  {Mitani}, \citenamefont {Hayashi}, \citenamefont {Yamanouchi},\ and\
  \citenamefont {Ohno}}]{torrejon14}%
  \BibitemOpen
  \bibfield  {author} {\bibinfo {author} {\bibfnamefont {J.}~\bibnamefont
  {Torrejon}}, \bibinfo {author} {\bibfnamefont {J.}~\bibnamefont {Kim}},
  \bibinfo {author} {\bibfnamefont {J.}~\bibnamefont {Sinha}}, \bibinfo
  {author} {\bibfnamefont {S.}~\bibnamefont {Mitani}}, \bibinfo {author}
  {\bibfnamefont {M.}~\bibnamefont {Hayashi}}, \bibinfo {author} {\bibfnamefont
  {M.}~\bibnamefont {Yamanouchi}}, \ and\ \bibinfo {author} {\bibfnamefont
  {H.}~\bibnamefont {Ohno}},\ }\bibfield  {title} {\enquote {\bibinfo {title}
  {Interface control of the magnetic chirality in {Co}{Fe}{B}/{Mg}{O}
  heterostructures with heavy-metal underlayers},}\ }\href@noop {} {\bibfield
  {journal} {\bibinfo  {journal} {Nat. Commun.}\ }\textbf {\bibinfo {volume}
  {5}},\ \bibinfo {pages} {4655} (\bibinfo {year} {2014})}\BibitemShut
  {NoStop}%
\bibitem [{\citenamefont {Miron}\ \emph {et~al.}(2011)\citenamefont {Miron},
  \citenamefont {Garello}, \citenamefont {Gaudin}, \citenamefont {Zermatten},
  \citenamefont {Costache}, \citenamefont {Auffret}, \citenamefont {Bandiera},
  \citenamefont {Rodmacq}, \citenamefont {Schuhl},\ and\ \citenamefont
  {Gambardella}}]{miron11}%
  \BibitemOpen
  \bibfield  {author} {\bibinfo {author} {\bibfnamefont {I.~M.}\ \bibnamefont
  {Miron}}, \bibinfo {author} {\bibfnamefont {K.}~\bibnamefont {Garello}},
  \bibinfo {author} {\bibfnamefont {G.}~\bibnamefont {Gaudin}}, \bibinfo
  {author} {\bibfnamefont {P.-J.}\ \bibnamefont {Zermatten}}, \bibinfo {author}
  {\bibfnamefont {M.~V.}\ \bibnamefont {Costache}}, \bibinfo {author}
  {\bibfnamefont {S.}~\bibnamefont {Auffret}}, \bibinfo {author} {\bibfnamefont
  {S.}~\bibnamefont {Bandiera}}, \bibinfo {author} {\bibfnamefont
  {B.}~\bibnamefont {Rodmacq}}, \bibinfo {author} {\bibfnamefont
  {A.}~\bibnamefont {Schuhl}}, \ and\ \bibinfo {author} {\bibfnamefont
  {P.}~\bibnamefont {Gambardella}},\ }\bibfield  {title} {\enquote {\bibinfo
  {title} {Perpendicular switching of a single ferromagnetic layer induced by
  in-plane current injection},}\ }\href@noop {} {\bibfield  {journal} {\bibinfo
   {journal} {Nature}\ }\textbf {\bibinfo {volume} {476}},\ \bibinfo {pages}
  {189} (\bibinfo {year} {2011})}\BibitemShut {NoStop}%
\bibitem [{\citenamefont {Kim}\ \emph {et~al.}(2012)\citenamefont {Kim},
  \citenamefont {Seo}, \citenamefont {Ryu}, \citenamefont {Lee},\ and\
  \citenamefont {Lee}}]{kim12}%
  \BibitemOpen
  \bibfield  {author} {\bibinfo {author} {\bibfnamefont {K.-W.}\ \bibnamefont
  {Kim}}, \bibinfo {author} {\bibfnamefont {S.-M.}\ \bibnamefont {Seo}},
  \bibinfo {author} {\bibfnamefont {J.}~\bibnamefont {Ryu}}, \bibinfo {author}
  {\bibfnamefont {K.-J.}\ \bibnamefont {Lee}}, \ and\ \bibinfo {author}
  {\bibfnamefont {H.-W.}\ \bibnamefont {Lee}},\ }\bibfield  {title} {\enquote
  {\bibinfo {title} {Magnetization dynamics induced by in-plane currents in
  ultrathin magnetic nanostructures with {Rashba} spin-orbit coupling},}\
  }\href@noop {} {\bibfield  {journal} {\bibinfo  {journal} {Phys. Rev. B}\
  }\textbf {\bibinfo {volume} {85}},\ \bibinfo {pages} {180404} (\bibinfo
  {year} {2012})}\BibitemShut {NoStop}%
\bibitem [{\citenamefont {Haney}\ \emph
  {et~al.}(2013{\natexlab{a}})\citenamefont {Haney}, \citenamefont {Lee},
  \citenamefont {Lee}, \citenamefont {Manchon},\ and\ \citenamefont
  {Stiles}}]{haney13}%
  \BibitemOpen
  \bibfield  {author} {\bibinfo {author} {\bibfnamefont {P.~M.}\ \bibnamefont
  {Haney}}, \bibinfo {author} {\bibfnamefont {H.-W.}\ \bibnamefont {Lee}},
  \bibinfo {author} {\bibfnamefont {K.-J.}\ \bibnamefont {Lee}}, \bibinfo
  {author} {\bibfnamefont {A.}~\bibnamefont {Manchon}}, \ and\ \bibinfo
  {author} {\bibfnamefont {M.~D.}\ \bibnamefont {Stiles}},\ }\bibfield  {title}
  {\enquote {\bibinfo {title} {Current induced torques and interfacial
  spin-orbit coupling: {Semiclassical} modeling},}\ }\href@noop {} {\bibfield
  {journal} {\bibinfo  {journal} {Phys. Rev. B}\ }\textbf {\bibinfo {volume}
  {87}},\ \bibinfo {pages} {174411} (\bibinfo {year}
  {2013}{\natexlab{a}})}\BibitemShut {NoStop}%
\bibitem [{\citenamefont {Haney}\ \emph
  {et~al.}(2013{\natexlab{b}})\citenamefont {Haney}, \citenamefont {Lee},
  \citenamefont {Lee}, \citenamefont {Manchon},\ and\ \citenamefont
  {Stiles}}]{haney13a}%
  \BibitemOpen
  \bibfield  {author} {\bibinfo {author} {\bibfnamefont {P.~M.}\ \bibnamefont
  {Haney}}, \bibinfo {author} {\bibfnamefont {H.-W.}\ \bibnamefont {Lee}},
  \bibinfo {author} {\bibfnamefont {K.-J.}\ \bibnamefont {Lee}}, \bibinfo
  {author} {\bibfnamefont {A.}~\bibnamefont {Manchon}}, \ and\ \bibinfo
  {author} {\bibfnamefont {M.~D.}\ \bibnamefont {Stiles}},\ }\bibfield  {title}
  {\enquote {\bibinfo {title} {Current-induced torques and interfacial
  spin-orbit coupling},}\ }\href@noop {} {\bibfield  {journal} {\bibinfo
  {journal} {Phys. Rev. B}\ }\textbf {\bibinfo {volume} {88}},\ \bibinfo
  {pages} {214417} (\bibinfo {year} {2013}{\natexlab{b}})}\BibitemShut
  {NoStop}%
\bibitem [{\citenamefont {Jamali}\ \emph {et~al.}(2013)\citenamefont {Jamali},
  \citenamefont {Narayanapillai}, \citenamefont {Qiu}, \citenamefont {Loong},
  \citenamefont {Manchon},\ and\ \citenamefont {Yang}}]{jamali13}%
  \BibitemOpen
  \bibfield  {author} {\bibinfo {author} {\bibfnamefont {M.}~\bibnamefont
  {Jamali}}, \bibinfo {author} {\bibfnamefont {K.}~\bibnamefont
  {Narayanapillai}}, \bibinfo {author} {\bibfnamefont {X.}~\bibnamefont {Qiu}},
  \bibinfo {author} {\bibfnamefont {L.~M.}\ \bibnamefont {Loong}}, \bibinfo
  {author} {\bibfnamefont {A.}~\bibnamefont {Manchon}}, \ and\ \bibinfo
  {author} {\bibfnamefont {H.}~\bibnamefont {Yang}},\ }\bibfield  {title}
  {\enquote {\bibinfo {title} {Spin-{Orbit} {Torques} in {Co}/{Pd} {Multilayer}
  {Nanowires}},}\ }\href@noop {} {\bibfield  {journal} {\bibinfo  {journal}
  {Phys. Rev. Lett.}\ }\textbf {\bibinfo {volume} {111}},\ \bibinfo {pages}
  {246602} (\bibinfo {year} {2013})}\BibitemShut {NoStop}%
\bibitem [{\citenamefont {Yu}\ \emph {et~al.}(2014)\citenamefont {Yu},
  \citenamefont {Upadhyaya}, \citenamefont {Fan}, \citenamefont {Alzate},
  \citenamefont {Jiang}, \citenamefont {Wong}, \citenamefont {Takei},
  \citenamefont {Bender}, \citenamefont {Chang}, \citenamefont {Jiang},
  \citenamefont {Lang}, \citenamefont {Tang}, \citenamefont {Wang},
  \citenamefont {Tserkovnyak}, \citenamefont {Amiri},\ and\ \citenamefont
  {Wang}}]{yu14}%
  \BibitemOpen
  \bibfield  {author} {\bibinfo {author} {\bibfnamefont {G.}~\bibnamefont
  {Yu}}, \bibinfo {author} {\bibfnamefont {P.}~\bibnamefont {Upadhyaya}},
  \bibinfo {author} {\bibfnamefont {Y.}~\bibnamefont {Fan}}, \bibinfo {author}
  {\bibfnamefont {J.G.}\ \bibnamefont {Alzate}}, \bibinfo {author}
  {\bibfnamefont {W.}~\bibnamefont {Jiang}}, \bibinfo {author} {\bibfnamefont
  {K.~L.}\ \bibnamefont {Wong}}, \bibinfo {author} {\bibfnamefont
  {S.}~\bibnamefont {Takei}}, \bibinfo {author} {\bibfnamefont {S.~A.}\
  \bibnamefont {Bender}}, \bibinfo {author} {\bibfnamefont {L.-T.}\
  \bibnamefont {Chang}}, \bibinfo {author} {\bibfnamefont {Y.}~\bibnamefont
  {Jiang}}, \bibinfo {author} {\bibfnamefont {M.}~\bibnamefont {Lang}},
  \bibinfo {author} {\bibfnamefont {J.}~\bibnamefont {Tang}}, \bibinfo {author}
  {\bibfnamefont {Y.}~\bibnamefont {Wang}}, \bibinfo {author} {\bibfnamefont
  {Y.}~\bibnamefont {Tserkovnyak}}, \bibinfo {author} {\bibfnamefont {P.~K.}\
  \bibnamefont {Amiri}}, \ and\ \bibinfo {author} {\bibfnamefont {K.~L.}\
  \bibnamefont {Wang}},\ }\bibfield  {title} {\enquote {\bibinfo {title}
  {Switching of perpendicular magnetization by spin-orbit torques in the
  absence of external magnetic fields},}\ }\href@noop {} {\bibfield  {journal}
  {\bibinfo  {journal} {Nat. Nanotech.}\ }\textbf {\bibinfo {volume} {9}},\
  \bibinfo {pages} {548} (\bibinfo {year} {2014})}\BibitemShut {NoStop}%
\bibitem [{\citenamefont {Pauyac}\ \emph {et~al.}(2013)\citenamefont {Pauyac},
  \citenamefont {Wang}, \citenamefont {Chshiev},\ and\ \citenamefont
  {Manchon}}]{pauyac13}%
  \BibitemOpen
  \bibfield  {author} {\bibinfo {author} {\bibfnamefont {C.~O.}\ \bibnamefont
  {Pauyac}}, \bibinfo {author} {\bibfnamefont {X.}~\bibnamefont {Wang}},
  \bibinfo {author} {\bibfnamefont {M.}~\bibnamefont {Chshiev}}, \ and\
  \bibinfo {author} {\bibfnamefont {A.}~\bibnamefont {Manchon}},\ }\bibfield
  {title} {\enquote {\bibinfo {title} {Angular dependence and symmetry of
  {Rashba} spin torque in ferromagnetic structures},}\ }\href@noop {}
  {\bibfield  {journal} {\bibinfo  {journal} {Appl. Phys. Lett.}\ }\textbf
  {\bibinfo {volume} {102}},\ \bibinfo {pages} {252403} (\bibinfo {year}
  {2013})}\BibitemShut {NoStop}%
\bibitem [{\citenamefont {Torrejon}\ \emph {et~al.}(2015)\citenamefont
  {Torrejon}, \citenamefont {Garcia-Sanchez}, \citenamefont {Taniguchi},
  \citenamefont {Shinha}, \citenamefont {Mitani}, \citenamefont {Kim},\ and\
  \citenamefont {Hayashi}}]{torrejon15PRB}%
  \BibitemOpen
  \bibfield  {author} {\bibinfo {author} {\bibfnamefont {J.}~\bibnamefont
  {Torrejon}}, \bibinfo {author} {\bibfnamefont {F.}~\bibnamefont
  {Garcia-Sanchez}}, \bibinfo {author} {\bibfnamefont {T.}~\bibnamefont
  {Taniguchi}}, \bibinfo {author} {\bibfnamefont {J.}~\bibnamefont {Shinha}},
  \bibinfo {author} {\bibfnamefont {S.}~\bibnamefont {Mitani}}, \bibinfo
  {author} {\bibfnamefont {J.-V.}\ \bibnamefont {Kim}}, \ and\ \bibinfo
  {author} {\bibfnamefont {M.}~\bibnamefont {Hayashi}},\ }\bibfield  {title}
  {\enquote {\bibinfo {title} {Current-driven asymmetric magnetization
  switching in perpendicularly magnetized {Co}{Fe}{B}/{Mg}{O}
  heterostructures},}\ }\href@noop {} {\bibfield  {journal} {\bibinfo
  {journal} {Phys. Rev. B}\ }\textbf {\bibinfo {volume} {91}},\ \bibinfo
  {pages} {214434} (\bibinfo {year} {2015})}\BibitemShut {NoStop}%
\bibitem [{\citenamefont {Liu}\ \emph {et~al.}(2012{\natexlab{c}})\citenamefont
  {Liu}, \citenamefont {Lee}, \citenamefont {Gudmundsen}, \citenamefont
  {Ralph},\ and\ \citenamefont {Buhrman}}]{liu12}%
  \BibitemOpen
  \bibfield  {author} {\bibinfo {author} {\bibfnamefont {L.}~\bibnamefont
  {Liu}}, \bibinfo {author} {\bibfnamefont {O.~J.}\ \bibnamefont {Lee}},
  \bibinfo {author} {\bibfnamefont {T.~J.}\ \bibnamefont {Gudmundsen}},
  \bibinfo {author} {\bibfnamefont {D.~C.}\ \bibnamefont {Ralph}}, \ and\
  \bibinfo {author} {\bibfnamefont {R.~A.}\ \bibnamefont {Buhrman}},\
  }\bibfield  {title} {\enquote {\bibinfo {title} {Current-{Induced}
  {Switching} of {Perpendicularly} {Magnetized} {Magnetic} {Layers} {Using}
  {Spin} {Torque} from the {Spin} {Hall} {Effect}},}\ }\href@noop {} {\bibfield
   {journal} {\bibinfo  {journal} {Phys. Rev. Lett.}\ }\textbf {\bibinfo
  {volume} {109}},\ \bibinfo {pages} {096602} (\bibinfo {year}
  {2012}{\natexlab{c}})}\BibitemShut {NoStop}%
\bibitem [{you()}]{you14}%
  \BibitemOpen
  \href@noop {} {}\bibinfo {note} {L. You, O. Lee, D. Bhowmik, D. Labanowski,
  J. Hong, J. Bokor, and S. Salahuddin, arXiv:1409.0620.}\BibitemShut {Stop}%
\bibitem [{com({\natexlab{a}})}]{comment_yan_arXiv}%
  \BibitemOpen
  \href@noop {} {} ({\natexlab{a}}),\ \bibinfo {note} {a similar approach was
  recently developed by Yan and Bazaliy, Phys. Rev. B $\bm{91}$ 214424
  (2015).}\BibitemShut {Stop}%
\bibitem [{com({\natexlab{b}})}]{comment1}%
  \BibitemOpen
  \href@noop {} {} ({\natexlab{b}}),\ \bibinfo {note} {the analytical formula
  of the critical current corresponding the case of $\beta=0$ is discussed in
  Ref. \cite{lee13}. The formula of Ref. \cite{lee13} is applicable to a large
  damping limit ($\alpha>0.03$), while we are interested in a low damping
  limit.}\BibitemShut {Stop}%
\bibitem [{\citenamefont {Bertotti}\ \emph {et~al.}(2009)\citenamefont
  {Bertotti}, \citenamefont {Mayergoyz},\ and\ \citenamefont
  {Serpico}}]{bertotti09}%
  \BibitemOpen
  \bibfield  {author} {\bibinfo {author} {\bibfnamefont {G.}~\bibnamefont
  {Bertotti}}, \bibinfo {author} {\bibfnamefont {I.}~\bibnamefont {Mayergoyz}},
  \ and\ \bibinfo {author} {\bibfnamefont {C.}~\bibnamefont {Serpico}},\
  }\href@noop {} {\emph {\bibinfo {title} {Nonlinear magnetization Dynamics in
  Nanosystems}}}\ (\bibinfo  {publisher} {Elsevier, Oxford},\ \bibinfo {year}
  {2009})\BibitemShut {NoStop}%
\bibitem [{\citenamefont {Wiggins}(2003)}]{wiggins03}%
  \BibitemOpen
  \bibfield  {author} {\bibinfo {author} {\bibfnamefont {S.}~\bibnamefont
  {Wiggins}},\ }\enquote {\bibinfo {title} {Introduction to applied nonlinear
  dynamical systems and chaos},}\ \ (\bibinfo  {publisher} {Springer},\
  \bibinfo {year} {2003})\ Chap.~\bibinfo {chapter} {1}\BibitemShut {NoStop}%
\bibitem [{\citenamefont {Taniguchi}(2015)}]{taniguchi15}%
  \BibitemOpen
  \bibfield  {author} {\bibinfo {author} {\bibfnamefont {T.}~\bibnamefont
  {Taniguchi}},\ }\bibfield  {title} {\enquote {\bibinfo {title} {Nonlinear
  analysis of magnetization dynamics excited by spin {Hall} effect},}\
  }\href@noop {} {\bibfield  {journal} {\bibinfo  {journal} {Phys. Rev. B}\
  }\textbf {\bibinfo {volume} {91}},\ \bibinfo {pages} {104406} (\bibinfo
  {year} {2015})}\BibitemShut {NoStop}%
\end{thebibliography}


\end{document}